\documentclass[article,twocolumn,floats,nofootinbib,nobibnotes,superscriptaddress,10pt]{revtex4-1}

\usepackage{graphicx,amsmath,amssymb,hyperref}
\hypersetup{colorlinks=true, linkcolor=red, citecolor=magenta, urlcolor=blue}
\usepackage{dcolumn}
\usepackage{afterpage}
\usepackage{xcolor,colortbl}
\usepackage{enumitem}
\usepackage{eqnarray}
\usepackage{makecell}
\usepackage{cancel}
\usepackage{etoolbox}
\usepackage{comment}
\usepackage{amsfonts,mathtools}
\usepackage{tabularx,booktabs}
\usepackage{multirow}
\usepackage{rotating}
\usepackage{array}
\newcolumntype{P}[1]{>{\centering\arraybackslash}p{#1}}
\usepackage{makecell}
\usepackage{refcount}

\usepackage{ulem} 
\usepackage{tikz}
\usepackage{bbding}
\usepackage{cancel}

\usepackage{pifont}

\newcommand{\Dttone}{\Delta^2_{T_{21}}}

\newcommand{\sigPS}{\sigma_{\rm PS}}
\newcommand{\tauPS}{\tau_{\rm PS}}
\newcommand{\sigmax}{\sigma_{\rm PS}^{\rm max}}
\newcommand{\dtau}{d\log_{10}\tau_{\rm PS}/d\log_{10}M_h}
\newcommand{\sigxsq}{\sigma_x^{2}}
\newcommand{\Mh}{M_h}
\newcommand{\Msun}{M_\odot}
\newcommand{\Mpcinv}{\,{\rm cMpc}^{-1}}

\newcommand{\zWF}{z_{\rm WF}}

\makeatletter
\newcommand\footnoteref[1]{\protected@xdef\@thefnmark{\ref{#1}}\@footnotemark}
\makeatother

\newcommand\affBS{  
\affiliation{Department of Physics, Ben-Gurion University of the Negev, Be'er Sheva 84105, Israel}
}

\newcommand\affEV{ 
\affiliation{School of Physics and Astronomy, Tel-Aviv University, Tel-Aviv 69978, Israel}
\affiliation{Department of Physics, Ben-Gurion University of the Negev, Be'er Sheva 84105, Israel}
}

\newcommand\affJM{
\affiliation{Department of Astronomy, The University of Texas at Austin,
2515 Speedway, Stop C1400, Austin, Texas 78712, USA}
\affiliation{Cosmic Frontier Center, The University of Texas at Austin, Austin, TX 78712}
}

\begin{document}

\title{When galaxies burst II. Implications of enhanced burstiness \\ for the 21-cm Cosmic Dawn signal
}

\author{Hovav Lazare}
\affBS
\author{Sarah Libanore}
\affBS
\author{Eleonora Vanzan}
\affEV
\author{Julian B. Mu\~noz}
\affJM
\author{Ely D. Kovetz}
\affBS

\begin{abstract}
Recent JWST observations suggest that star formation in the early universe was substantially burstier than assumed in standard models. 
Such burstiness can be described as a stochastic process characterized by the burst amplitude and the coherence time of star formation epochs.
In this paper, we investigate how bursty star formation modifies the 21-cm power spectrum during Cosmic Dawn through its impact on the non-local radiation fields that govern its evolution, namely Lyman-$\alpha$ and X-ray backgrounds. To do so, we introduce an unequal-time correlation in the star-formation-rate density sourcing the two fields and we compute its impact using the analytical framework implemented in the public code \texttt{Zeus21}. We find that the burstiness-induced time correlation produces a shot-noise-like contribution in the Lyman-$\alpha$ and X-ray fields, enhancing both their auto- and cross-power spectra while leaving the global 21-cm signal, $T_{21}(z)$, unchanged.
As a result, the 21-cm power spectrum is strongly modified by a shot-noise-like contribution at the beginning of the Cosmic Dawn, where the signal is dominated by lower-mass halos ($\Mh\lesssim10^{10}\,\Msun$), and is boosted by a factor of a few near the Wouthuysen-Field absorption trough. Elsewhere at low redshift, where the clustering signal dominates and larger halos drive the signal, the burstiness component is negligible.

\end{abstract}

\maketitle

\section{Introduction}\label{sec:intro}

Since its launch, the James Webb Space Telescope (JWST) has revolutionized our view of galaxies in the early Universe. Observations at $z\gtrsim9$ have revealed an ultraviolet (UV) luminosity function that exceeds the predictions of standard galaxy-formation models~\cite{Mason:2022tiy,Harikane:2022rqt,Finkelstein:2024,Donnan:2024,Robertson:2023nyv}, uncovering tensions that suggest the physical processes governing high-redshift galaxies may differ substantially from those inferred at lower redshift. A leading explanation is that star formation in these systems is significantly more bursty than commonly assumed~\cite{Robertson:2022gdk,Mirocha:2022,Sun:2023ocn,Pallottini:2023yqg}.

Indeed, a growing body of observational evidence points toward highly bursty star-formation histories (SFHs) in high-redshift galaxies, characterized by episodes of enhanced star formation occurring on timescales of order tens of Myr. Such behavior is expected in feedback-regulated galaxies, where star formation fluctuates stochastically on timescales that are short compared to the Hubble time but long compared to the local free-fall time~\cite{Sparre:2017,Faucher-Giguere:2017lgp,Furlanetto:2022,Sun:2023}. Further evidence for this picture comes from the broad distribution of the H$\alpha$-to-UV luminosity ratio~\cite{Clarke:2024,Endsley:2024,Sun:2025cpl,Munoz:2026gxp}. Since H$\alpha$ emission traces star formation over timescales of a few Myr, whereas the UV continuum is sensitive to star formation averaged over $\sim100\,\mathrm{Myr}$, bursty star-formation histories naturally produce significant scatter in this ratio. 

Recently, Ref.~\cite{Munoz:2026gxp} (hereafter M26) developed a quantitative framework for describing bursty star formation, calibrated against a combination of UV luminosity function, clustering, and H$\alpha$/UV observations. Their model characterizes burstiness through the root mean square (rms) of the scatter of the log-star formation rate (SFR), $\sigPS$ (which quantifies the burst amplitude), its dependence on halo mass, $d\sigPS/d\log_{10}\Mh$, and the characteristic coherence time between bursts, $\tauPS$. M26 found that burstiness increases toward lower halo masses, implying that stochastic star formation is a common feature of the high-$z$ galaxy population.

The implications of this picture extend well beyond galaxy surveys. In a companion paper~\cite{Kovetz:2026xfs}, we propagated the M26 framework to line-intensity mapping (LIM,~\cite{Kovetz:2017agg,Bernal:2022jap,Chang:2026ake}) observables and showed that bursty star formation enhances the shot-noise component of LIM power spectra. The magnitude of this enhancement depends on both $\tauPS$ and the characteristic stellar ages traced by the target emission line. Because LIM is sensitive to the cumulative emission of faint galaxies residing in low-mass halos, it provides a particularly powerful probe of burstiness, as Ref.~\cite{Liu:2024fti} also pointed out in the context of [CII] LIM forecasts.

The increased burstiness in faint galaxies primarily hosted by low-mass dark matter (DM) halos, suggests that similar effects may be even more pronounced in earlier epochs, including the Epoch of Reionization (EoR) and Cosmic Dawn. These eras are most directly accessible through the 21-cm signal~\cite{LoebFurlanetto:2013,
Pritchard:2011xb,Furlanetto:2006jb, Madau:1996cs, Loeb:2003ya, Pober:2013jna, Liu:2022iyy}, which provides a uniquely rich probe of the first billion years of cosmic history. The increased variability and efficiency of high-$z$ star formation leave significant imprints on the 21-cm power spectrum, with signatures that will soon become detectable by current and upcoming interferometers and global signal experiments~\cite{Libanore:2023oxf,Hassan:2023asd,Dhandha:2025gib}.

Unlike other LIM tracers, the 21-cm signal is governed by non-local radiation backgrounds and therefore connects to galaxy formation through a fundamentally different mechanism.
The evolution of the 21-cm signal and its fluctuations is driven by the interplay between the properties of neutral hydrogen (HI) in the intergalactic medium (IGM) and the radiation emitted by the first galaxies. Two radiation fields are particularly important for modeling the 21-cm signal during Cosmic Dawn. The first is the Lyman-$\alpha$ (Ly$\alpha$) background, which couples the spin temperature of the gas---set by the relative occupation of the singlet and triplet hyperfine states---to the kinetic temperature through the Wouthuysen-Field (WF) effect~\cite{Wouthuysen:1952,Field:1959,Hirata:2005mz,Pritchard:2005an}. The second is the X-ray background, produced primarily by early black holes and high-mass X-ray binaries, which heats the IGM at $z\lesssim 20$~\cite{Madau:2016jbv,Pritchard:2006sq,Mesinger:2012ys,Pacucci:2014wwa,Fialkov:2014kta}. At later times, as the Universe approaches the EoR, the ionizing radiation field becomes an additional key driver of the 21-cm signal. In this work, we focus on the high-redshift regime, $z \gtrsim 10$, where the ionized fraction remains small and reionization effects can be neglected to first approximation.

Both the Ly$\alpha$ and X-ray fields act in a non-local manner, since their photons typically travel large distances from their sources before being absorbed. As a result, the 21-cm brightness temperature at a given position and redshift depends on the past star-formation rate density (SFRD) in surrounding regions. This introduces an effective {\it retarded response} of the 21-cm signal to the SFRD, mediated by the propagation of Ly$\alpha$ and X-ray photons. Consequently, the relevant fluctuations are intrinsically time dependent and cannot be fully described by equal-time correlators alone, motivating the use of an unequal-time formalism. Unequal-time correlations have been employed in galaxy surveys that measure sufficiently large radial separations that the density field evolves appreciably, and where redshift-space distortions and general relativistic projection effects may also introduce additional time dependence~\cite{Kitching:2016xcl,Chisari:2019tig,delaBella:2020rpq,Raccanelli:2023fle,Spezzati:2025ntb}. In the context of 21-cm science, particularly during the EoR, different summary statistics have been proposed to account for the rapid evolution of the underlying fluctuation field~\cite{Datta:2006vh,Pramanick:2025byb,Blamart:2025jyt}. 
Unlike these applications, driven by either the deterministic evolution or the geometric projection of large-scale structure over cosmological distances, the burstiness we consider in this work affects the SFR on much shorter, intrinsically stochastic timescales, set by individual halos rather than by the growth of cosmic structure. This is conceptually closer to the case of quasars studied in Ref.~\cite{Meiksin:2018wij}, which developed a time-dependent fluctuation formalism to account for finite quasar lifetimes relative to the long photon propagation times governing the photoionizing background. They showed that unequal-time correlations are necessary to correctly model the resulting shot-noise, since the contribution from discrete sources with finite lifetimes is intrinsically coupled to the propagation kernel of the field they generate. 

In this work, we perform an analogous convolution between the stochastic, finite-coherence-time star-formation histories of individual halos and the propagation kernels of the Ly$\alpha$ and X-ray photons sourcing the 21-cm signal.
We incorporate burstiness into our analytical framework as a non-local mapping between the per-halo SFR variance and the fluctuations in the radiation fields. This involves an integral over radial shells around the point in which the signal is computed, which leads to an effective {\it unequal-time cross-correlation} in the SFRD across different shells. As we show throughout the text, this introduces a boosted shot-noise-like contribution to the Ly$\alpha$ and X-ray fields. 

Other 21-cm codes, such as Ref.~\cite{Reis:2021sqh}, or the recent version of the semi-numerical code \texttt{21cmFAST}~\cite{Davies:2025wsa} (see Refs.~\cite{Mesinger:2010ne,Murray:2020trn,Munoz:2021psm} for earlier versions), as well as the semi-analytic galaxy formation model \texttt{Meraxes}~\cite{Poole:2015mhx, Mutch:2016, Balu:2022dtp} which incorporates a modified \texttt{21cmFAST} reionization module, have also accounted for a similar effect, by introducing discrete halos with astrophysical properties scattered around the mean relation and correlated in time (\texttt{21cmFAST}), or by accounting for it implicitly through the galaxy formation model (\texttt{Meraxes}). The above codes implement a semi-numerical simulation of the evolution of the 21cm signal, which can last from a couple of minutes  up to a few days. 
Here, we implement this effect within the public, fast ($\mathcal{O}(1\, \rm sec)$), analytic code \texttt{Zeus21}~\cite{Munoz:2023kkg,Cruz:2024fsv},\footnote{\url{https://github.com/ZeusCosmo/Zeus21}} extending its standard formalism to include both the deterministic shot noise associated with source discreteness and the additional contribution arising from bursty star formation. We then study the impact of this enhanced stochasticity on the global 21-cm signal and its power spectrum.
The mean signal remains unchanged given the conventions chosen in this work, but the power spectrum is significantly modified in specific regimes. (i) During Cosmic Dawn, the 21-cm power spectrum is shaped by a near-cancellation between Ly$\alpha$ coupling and X-ray heating at $z\sim 15-17$, producing a characteristic near-null; the burst-induced shot noise is positive-definite and generically fills in this suppression, enhancing the power spectrum rather than preserving the cancellation.
(ii) A second regime of strong impact occurs at very high redshift, $z\gtrsim 20$. In this regime, the signal is dominated by low-mass halos with $\Mh \sim 10^{9}\,\Msun$, which are poorly constrained and require an extrapolation of the M26 model. We explore several truncation schemes for this low-mass extrapolation and assess their impact, highlighting the associated theoretical uncertainty. In all cases, we assume the evolution of the 21-cm signal is driven by Population~II (PopII) stars only, which in our model form in atomic-cooling galaxies (ACGs). Extending our formalism to Population~III (PopIII) stars, expected to form in molecular-cooling galaxies (MCGs) hosted by mini-halos with mass $\lesssim10^8M_\odot$, is in principle straightforward. However, the burstiness properties of MCGs are currently unconstrained by data~\cite{Venditti:2025mgi}; if these systems are bursty, their properties could differ substantially from those of ACGs, and should instead be anchored in physically motivated star-formation scenarios rather than in the data-calibrated model we adopt here. For this reason, we do not address MCGs and leave their inclusion to future work.

Our  results show that including burstiness in the modeling can be significant in specific regimes of the 21-cm signal evolution and enables a reliable and fast interpretation of forthcoming observations with low-frequency interferometers such as HERA~\cite{DeBoer:2016tnn,HERA:2021bsv,HERA:2022wmy} and LOFAR~\cite{LOFAR:2013jil,
Mertens:2020llj,Ceccotti:2025bcd}. 
 
The rest of this paper is structured as follows. Section~\ref{sec:bursty} introduces the burstiness model, summarizing the M26 formalism. Section~\ref{sec:shotnoise} presents our key result: we develop the formalism to compute the SFRD power spectrum by summing over contributions of discrete halos. This allows us to define the shot-noise component of the power spectrum and to incorporate the effects of burstiness through an unequal-time correlation between SFHs.
Section~\ref{sec:shells_fluxes} propagates these results to the Ly$\alpha$ and X-ray fields required for computing the 21-cm power spectrum, and characterizes the relative importance of burstiness as a function of redshift. Section~\ref{sec:results} presents our results for the 21-cm power spectrum: Figure~\ref{fig:ratio_map} is the central result of this paper, quantifying the impact of burstiness across redshift and scales.
Finally, Sec.~\ref{sec:Discussion} discusses the dependence of our results on different parameters, the limitations of our analysis, and its possible extensions. We present out conclusions in Section~\ref{Sec:conclusions}.


\section{Burstiness Effective Model}\label{sec:bursty}

To model burstiness in the SFR of high-$z$ galaxies, we adopt an effective model built in analogy to M26. We assume that the growth of galaxies over time $t$, i.e.,~their SFH $(\dot{M}_*)$, is described as a stochastic process that can be decomposed into the product of an average component with lognormal fluctuations~\cite{Pallottini:2023yqg}. 
In this context, we can express the SFR of a DM halo of mass $M_h$ at time $t$ as
\begin{equation}
    \dot{M}_*(M_h,t) \;=\; \overline{\dot{M}}_*(M_h, t)\;e^{x(t) - \sigma_x^2(M_h)/2}\,,
\label{eq:SFR_lognormal}
\end{equation}
where $\overline{\dot{M}}_*(M_h, t)$ is the mean SFH, averaged over many bursts, and $x(t)$ is the zero-mean Gaussian process that modulates the SFR of each individual galaxy.\footnote{Our convention is different from M26: while we anchor the SFH amplitude on the mean by subtracting $\sigma_x^2/2$, M26 defined 
\[\dot{M}_*(M_h,t) \;=\; \overline{\dot{M}}_*(M_h, t)\;e^{x(t)},\] and anchored it to the median; in their case, the mean of the SFH is $\overline{\dot{M}}_*(M_h, t)\,e^{\sigma_x^2/2}$. One can easily transform one prescription into the other by reparametrizing the amplitude of the mean SFR. 
By using the mean-anchored convention, we
ensure that $\overline{\dot{M}}_*(M_h, t)$ is directly calibrated on data. Propagating this into our pipeline leaves the average SFRD unchanged and does not modify the amplitude of the 21-cm global signal (neglecting fluctuation-induced corrections to a nonlinear sky average) targeted by EDGES~\cite{Bowman:2018yin}/ SARAS~\cite{Singh:2021mxo}, but only its power spectrum.} 
While 
$\overline{\dot{M}}_*(M_h,t)$ can be identified with the deterministic SFR prescription at $z = z(t)$ (presented in  Appendix~\ref{sec:Zeus21}), we treat $x(t)$ as a stationary Ornstein--Uhlenbeck (OU)~\cite{Uhlenbeck:1930zz} process,~i.e.,~a Gaussian process with autocorrelation
\begin{equation}
\xi_x(\Delta t) := \langle x(t)\,x(t')\rangle \;=\; \frac{\sigPS^2(M_h)}{2}\,e^{-|t-t'|/\tau_{\rm PS}}\,.
\label{eq:xi_x}
\end{equation}
Here, $\sigPS$ and $\tauPS$ parametrize the strength and coherence timescale of the bursts respectively~\cite{Caplar:2019,Bouche:2009gm,Dave:2012}. We note that by setting $\Delta t=0$, we recover the variance of $x(t)$, namely $\sigma_x^2(M_h)=\sigma_{\rm PS}^2/2$. For future use, we also define
\begin{equation}\label{eq:y_t}
    y(M_h,t)=\frac{\dot{M}_*(M_h,t)}{\overline{\dot{M}}_*(M_h,t)}=e^{x(t)-\sigma_x^2(M_h)/2};
\end{equation}
in our mean-anchored convention, this lognormal variable has mean $\langle y(t)\rangle=1$ and two-point function 
\begin{equation}\label{eq:y_autocorr}
    \xi_y(\Delta t):=\langle y(t)\,y(t')\rangle \;=\; {\rm exp}\left[\sigma_x^2(M_h)\,e^{-|t-t'|/\tau_{\rm PS}}\right]\,,
\end{equation}
which depends on the same parameters as $x(t)$. The interested reader can find more details in Appendix~\ref{app:detailed_form}.

In general, $\sigma_{\rm PS}$ and $\tauPS$ can depend on the mass of the host DM-halo; we parameterize this possibility via
\begin{align}
\sigPS(M_h) \;=\; \sigPS +
\frac{d\sigPS}{d\log_{10}\!M_h}\bigl(\log_{10}\!M_h - \log_{10}\!M_h^{\rm piv}\bigr),
\label{eq:sigPS_M}
\end{align}
and
\begin{align}
\tauPS(M_h) \;=\; \tauPS +
\frac{d\tauPS}{d\log_{10}\!M_h}\bigl(\log_{10}\!M_h - \log_{10}\!M_h^{\rm piv}\bigr).
\label{eq:tauPS_M}
\end{align}

To calibrate $\sigPS, \tauPS$ and their halo-mass slopes, M26 anchors $M_h^{\rm piv}$ at $10^{10}M_\odot$, and simultaneously fits the UV luminosity function, H$\alpha$/UV ratio and clustering data observed by JWST+HST (Hubble Space Telescope) at $z\sim 4-6$. Their results, which we adopt as fiducial values in our analysis, are summarized in Tab.~\ref{tab:fid}. 
M26 did not find any significant evidence for variation of $\tauPS$ with halo mass, and the 68\% CL obtained stems from the Gaussian prior rather than the data. Thus, we set $\dtau \!=\!0$ as our fiducial value, and examine in Sec.~\ref{sec:tau_evol} whether the 21-cm power spectrum may be sensitive to its value and yield stronger constraints.
\begin{table}[t!]
\renewcommand{\arraystretch}{1.5}
    \centering
    \begin{tabular}{|c|c|}
    \hline
         Parameter & Fiducial \\
    \hline
$\sigma_{\rm PS}$ & ${\mathbf{ 2.1}}^{+0.11}_{-0.08}$\\
$\tau_{\rm PS}$ & ${\mathbf{ 25}}^{+27}_{-11}\,${\rm Myr}\\
$d\sigma_{\rm PS}/d\log_{10}M_h$ & ${\mathbf{ -0.50}}\pm 0.13$\\
$\dtau$ & ${\mathbf 0}\,(-0.04\pm 0.26)$\\
\hline 
$\sigma_{\rm PS}^{\rm max}$ & $\in \{{\mathbf{ 2.3}},3.5\}$ \\
    \hline
    \end{tabular}
    \caption{Fiducial values of the burstiness-related parameters adopted in our analysis. We rely on the results of M26, for which we also quote the uncertainties, except for $\sigmax$, for which we test 
    different values. If not differently specified, our baseline assumes $\sigma_{\rm PS}^{\rm max}=2.3$. We note that although Eq.~\eqref{eq:tauPS_M} uses a linear slope for $\tauPS$, in practice we use a logarithmic slope.}
    \label{tab:fid}
\end{table}

In a more general framework, the burstiness parameters may also depend on redshift; since M26 found no significant $z$-dependence within the calibration range, we neglect it in our analysis and use $z$-independent $\sigPS$ and $\tauPS$. This is a key and potentially vulnerable assumption in our work, as we apply the conclusions of M26 to a redshift regime distinct from that of the data they used.

\begin{figure}[t!]
\centering
\includegraphics[width=\columnwidth]{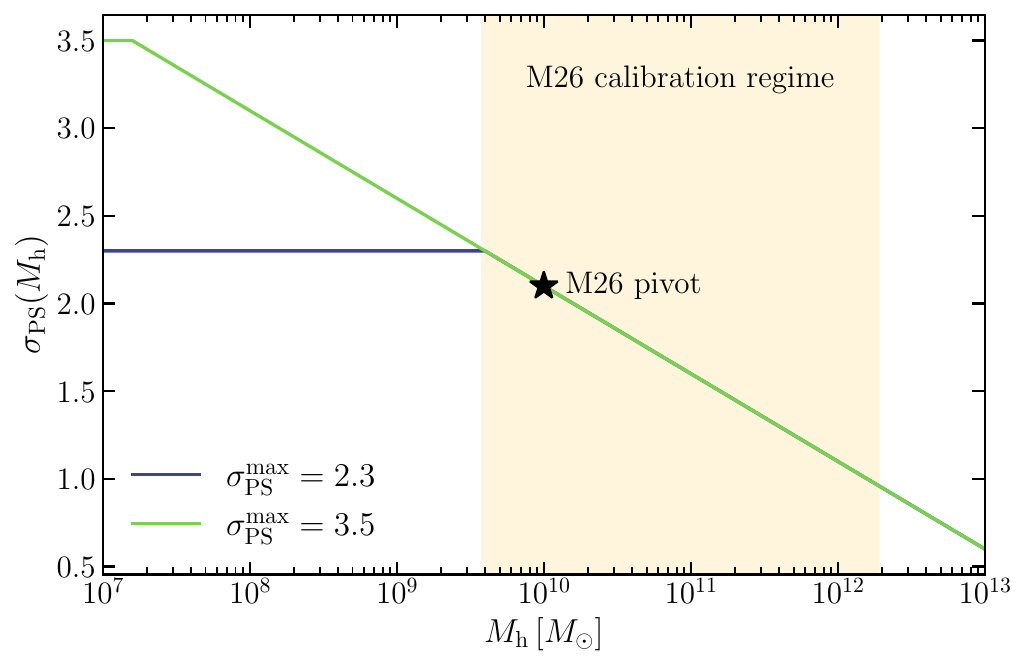}
\caption{Halo-mass dependence of the burstiness amplitude $\sigma_{\rm PS}(M_{\rm h})$ with amplitude $\sigPS$ at $M^{\rm piv}_{\rm h}$ and slope
  calibrated by M26 (see Tab.~\ref{tab:fid}). We clamp the evolution of $\sigma_{\rm PS}$ from above by two values of the saturation cap, $\sigma_{\rm PS}^{\rm
   max} = \{2.3, 3.5\}$. The first value corresponds to $\sigma_{\rm PS}$ at the lowest DM-halo mass probed by M26 data; the second, instead, allows us to consistently extrapolate $\sigma_{\rm PS}$ over all the DM-halo masses relevant for Cosmic Dawn 21-cm signal (see also Sec.~\ref{sec:Discussion} and Fig.~\ref{fig:kernel_Mh}). 
   The shaded band marks the M26 calibration regime in halo mass, estimated from the UV magnitude of the galaxies in their UV luminosity function and H$\alpha/UV$ datasets.}
\label{fig:sigma_Mh}
\end{figure}

Furthermore, as Fig.~\ref{fig:sigma_Mh} shows, our fiducial parameters introduce an increasing burstiness toward galaxies hosted by the low-mass DM halos which dominate at the high redshifts our work addresses. 
However, the data used in M26 only provide access to galaxies having UV absolute magnitude $M_{\rm UV}\lesssim -15$, typically hosted by halos with $M_h\gtrsim 5\times 10^9\,M_\odot$ (yellow, shaded area in Fig.~\ref{fig:sigma_Mh}). This limits the reliability of applying the M26 results to the faint systems required by our analysis. 
To account for this, we introduce the parameter $\sigma_{\rm PS}^{\rm max}$ as a cap to  $\sigma_{\rm PS}(M_h)$, preventing it from reaching unphysical values. As indicated in Tab.~\ref{tab:fid}, since the cap parameter is unconstrained by current observations, we test how our results are affected by different choices (conservative or aggressive) for its value.
We emphasize that this cap is best understood not as a hard physical ceiling, but as
the scale beyond which the M26 phenomenological extrapolation almost certainly 
ceases to be self-consistent; in Sec.~\ref{sec:Discussion}, we further comment about this.  

The effect of SFR burstiness on the UV luminosity function and on LIM power spectra has been studied in M26 and Ref.~\cite{Kovetz:2026xfs}, respectively. In the following sections, we complete the picture by estimating how burstiness impacts the 21-cm power spectrum, using the SFRD as the main building block for its computation.

\vspace*{-.3cm}
\section{Effects of burstiness on the star formation rate density}\label{sec:shotnoise}

The burstiness model described in Sec.~\ref{sec:bursty} introduces a temporal correlation in the evolution of the SFR within a single halo. In fact, the properties of the OU correlator in Eq.~\eqref{eq:xi_x} can be used to compute the correlation between any two times $t_{1}, t_{2}$ in the SFH of a DM-halo with mass $M_h$. As detailed in Appendix~\ref{app:detailed_form}, this results in
\begin{equation}
\begin{aligned}
\bigl\langle&\dot{M}_*(M_h, t_{1})\,\dot{M}_*(M_h, t_{2})\bigr\rangle
\;=\;\\
&=\overline{\dot{M}}_*(M_h, t_{1})\,\overline{\dot{M}}_*(M_h, t_{2}) \exp\!\Bigl[\sigma_x^{2}(M_h)\,e^{-|t_{1}-t_{2}|/\tau_{\rm PS}}\Bigr].
\label{eq:single_halo_master}
\end{aligned}
\end{equation}
This equation presents two useful limits: if $t_1=t_2=t$, the equation reduces to the computation of the SFR variance inside a single halo:
\begin{equation}
\bigl\langle\dot{M}_*^{2}(M_h, t)\bigr\rangle \;=\; \overline{\dot{M}}_*^{2}(M_h, t)\,e^{\sigma_x^{2}(M_h)},
\end{equation}
which can be thought of as a stochastic boost, due to variations in the SFR without any characteristic temporal correlation. Instead, if $|t_1-t_2|\!\gg\!\tau_{\rm PS}$, the OU correlator is exponentially small and the bracket in Eq.~\eqref{eq:single_halo_master} reduces to $\overline{\dot{M}}_*^{2}(M_h)$,~i.e.,~the deterministic limit. Between
these two limits, $\langle\dot{M}_*(M_h,t_1)\dot{M}_*(M_h,t_2)\rangle$ smoothly interpolates as a function of $|t_1-t_2|/\tau_{\rm PS}$.

In our formalism and in agreement with M26, the burstiness variable $x(t)$ is independent in each halo; thus, computing the correlation between SFHs of two different halos would return zero. Environmental effects,~e.g.~halo mergers or large-scale assembly bias, could change this picture and cause coherent bursts, temporally correlated, in neighboring halos. Although we briefly comment on this in Sec.~\ref{sec:extensions}, a detailed treatment of these scenarios is beyond the scope of this paper and left for future work.

Equation~\eqref{eq:single_halo_master} carries a crucial implication that forms the basis of the discussion in the following Sections: the burst-induced enhancement introduces an {\it unequal-time same-halo SFR autocorrelation effect}.

\subsection{Burstiness induced boosted shot noise}

To proceed with our computation, we first need to evaluate the SFRD and quantify the impact of burstiness on it. To do so, we adopt a framework that differs from the standard halo-model formalism~\cite{Cooray:2002dia}, which was employed, for example, in Refs.~\cite{Schaan:2021gzb,Schneider:2020xmf} for LIM and 21-cm studies. Instead, we treat each halo as a point source characterized by its instantaneous SFR, hence introducing the time coordinate $t$ among the variables of our model. 

In the burstiness scenario, the SFRD at time $t$ is:
\begin{equation}
\dot{\rho}_*(\boldsymbol{x}, t) \;=\; \sum_i \delta_D^{(3)}(\boldsymbol{x} - \boldsymbol{x}_i)\,\overline{\dot{M}}_*(M_h^i, t)\,y_i(t),
\label{eq:rhostar_setup}
\end{equation}
where each $i$-th halo is localized in $\boldsymbol{x}_i$ by the 3D Dirac delta $\delta_D^{(3)}$, and it contributes to the total SFRD through a deterministic mean SFR, set by its own mass $M_h^i$, and the multiplicative stochastic enhancement $y_i(t)$ introduced in Eq.~\eqref{eq:y_t}. 
We Fourier transform Eq.~\eqref{eq:rhostar_setup} and define the cross-power spectrum as 
\begin{equation}
\begin{aligned}
    \langle&\hat{\dot{\rho}}_*(\boldsymbol{k},t_1)\hat{\dot{\rho}}^\dagger_*(\boldsymbol{k}',t_2)\rangle=\delta^{(3)}_D(\boldsymbol{k}-\boldsymbol{k}')\\
    &\times\sum_{ij}\overline{\dot{M}}_*(M_h^i,t_1)\overline{\dot{M}}_*(M_h^j,t_2)\langle y_i(t_1)y_j(t_2)e^{-i\boldsymbol{k}\cdot(\boldsymbol{x}_i-\boldsymbol{x}_j)}\rangle
\label{eq:SFRD_2pt}
\end{aligned}
\end{equation}
where $\hat{.}$ indicates the Fourier transform, and $\dagger$ the complex conjugate. Crucially, Eq.~\eqref{eq:SFRD_2pt} accounts for correlations between different sources both in space and across cosmic time. The resulting unequal-time power spectrum naturally captures the time dependence of SFRD fluctuations, which is essential for modeling bursty star formation with finite temporal coherence. While, to our knowledge, this is the first application of the unequal-time formalism to burstiness in the context of the 21-cm signal, a similar approach (with different scope and derivation) has been employed in Ref.~\cite{Meiksin:2018wij}.  

Equation~\eqref{eq:SFRD_2pt} can be more easily interpreted by splitting the sum into the $i\neq j$ and $i=j$ contributions. In fact, the off-diagonal terms identify the two-point function between different halos, i.e.,~their clustering. As anticipated in the previous Section, this term is unchanged by burstiness in our model, due to the halo-independent stochasticity assumption. 
On the other hand, summing over the diagonal terms $i=j$ accounts for the contribution of the single-halo SFR autocorrelation at two different times, $t_1$ and $t_2$, namely of the quantity computed in Eq.~\eqref{eq:single_halo_master}. Moving from the discrete picture to a continuous setup using the halo mass function (HMF), we sum over halos of different masses with appropriate weighting, and obtain:
\begin{equation}
\begin{aligned}
&P^{\rm shot}_{\dot{\rho}_*}(k; t_1, t_2)\,=\,\\
&\quad\int\! dM_h\,\frac{dn(z_{\rm ref})}{dM_h}\,\overline{\dot{M}}_*(M_{h}(z_1), z_1)\overline{\dot{M}}_*(M_{h}(z_2), z_2)\\
&\quad\quad\times
\exp\!\Bigl[\sigma_x^{2}(M_{h}(z_{\rm ref}))\,e^{-|t_1\!-\!t_2|/\tau_{\rm PS}}\Bigr].
\end{aligned}
\label{eq:Pshot_master}
\end{equation}
The quantity in Eq.~\eqref{eq:Pshot_master} resembles a $k$-independent, one-halo contribution, and can therefore be interpreted as a white shot-noise term. Even in the absence of burstiness, i.e.~$\sigPS\!=\!0$, $P_{\dot{\rho}_*}^{\rm shot}(k)=\int dM_h dn/dM_h\overline{\dot{M}}_*^2(M_h)\neq 0$ may have a non-negligible impact. Throughout the rest of the paper, we will refer to this case as the {\it deterministic shot noise}. For clarity, we emphasize that our shot-noise term arises from a formalism that differs from the textbook treatment used in galaxy surveys, where the observable is inherently discrete. In our case, halo discreteness propagates into the construction of the continuous SFRD field. As we will show in Sec.~\ref{sec:shells_fluxes}, this has important consequences for the modeling of the radiation fields sourced by the SFRD, which propagate through the Universe and drive the evolution of the 21-cm power spectrum.

Before proceeding, let us further examine the properties of Eq.~\eqref{eq:Pshot_master}. We identify $z_{1,2}=z(t_{1,2})$ as the redshifts corresponding to the cosmic times $t_{1,2}$ at which the SFRD is evaluated. At each epoch, the SFRD is computed within individual halos. The halo mass evolves in time due to mass accretion, for which we adopt the exponential growth model $M_h(z)=M_{h,0}\exp(-\alpha z)$, where $\alpha=0.79$ as calibrated from simulations~\cite{Dekel:2013uaa}. We neglect scatter in individual halo accretion histories around the mean growth track; alternatively, such scatter can be absorbed into the scatter amplitude of the SFR.

Notably, we evaluate the halo mass function at $z_{\rm ref}=\max(z_1,z_2)$, thereby anchoring the halo abundance in each mass bin at the higher redshift and assuming that the HMF does not evolve over the characteristic timescale of the burst ($|t_1-t_2|\ll H^{-1}(z)$). This approximation neglects halo mergers, which would reduce the number of halos contributing coherently at both epochs in Eq.~\eqref{eq:Pshot_master}; accordingly, our formulation should be interpreted as an upper bound on the cross-redshift correlation strength.

We further assume that the displacement of individual halos over the burst timescale is negligible (sub-kpc compared to the Mpc scales of interest). Finally, the amplitude of the SFR correlation, $\sigma_x^2(M_h(z_{\rm ref}))$, is evaluated at the higher redshift, consistent with the OU process governing the burst statistics.

\section{Effects of burstiness on the 21-cm power spectrum}\label{sec:shells_fluxes}

During Cosmic Dawn, the evolution of the 21-cm brightness temperature $T_{21}$ is primarily driven by the radiation fields galaxies produce. In particular, Ly$\alpha$ photons couple the spin temperature of neutral hydrogen to the gas kinetic temperature through the WF effect; at the same time, X-ray radiation heats the IGM, which would otherwise cool adiabatically.
Both Ly$\alpha$ and X-ray photons can travel significant distances before being absorbed, making the 21-cm signal inherently non-local in its dependence on galaxy formation. 
As a result, computing the signal at a given position $\boldsymbol{x}$ requires accounting for the past SFR in surrounding regions. 

In practice, to compute the fluxes of the radiation fields that govern the 21cm signal,
we assume that the instantaneous luminosities of these fields are proportional to the SFR at the time of emission. This is a simplifying assumption we use to approximate the fluxes to first order, since these luminosities have a non-trivial dependence on the SFH (see M26 for luminosity computation including SFHs for specific lines, and Sec.~\ref{sec:Discussion} for further discussion). We leave a more generalized formalism to future studies. 
Within this framework, the relevant fluxes $J_{{\rm Ly}\alpha,T_X}(\boldsymbol{x})$ can be expressed as weighted averages of the SFRD over comoving distances $R$ along the past lightcone~\cite{Pritchard:2006sq},
\begin{equation}
    J_{{\rm Ly\alpha},T_X}\propto\int dR\, c_{\alpha,X}(R)\dot{\rho}_*(R),
\label{eq:fluxes_general}
\end{equation}
where the coefficients $c_{\alpha,X}$ account for photon propagation, astrophysical and cosmological properties. 

This non-locality becomes particularly important in the presence of bursty star formation. Since the Ly$\alpha$ and X-ray radiation fields are obtained through integrals over the past lightcone, the 21-cm signal depends on radiation emitted by individual halos at different times. The impact of Poisson fluctuations on the Ly$\alpha$ power spectrum was first described in Ref.~\cite{Barkana:2004vb} and then Ref.~\cite{Reis:2020hrw}. On top of this, in the presence of burstiness, these fields inherit the unequal-time correlations introduced in Sec.~\ref{sec:shotnoise}.
\begin{figure}[ht!]
    \centering
    \includegraphics[width=.8\linewidth]{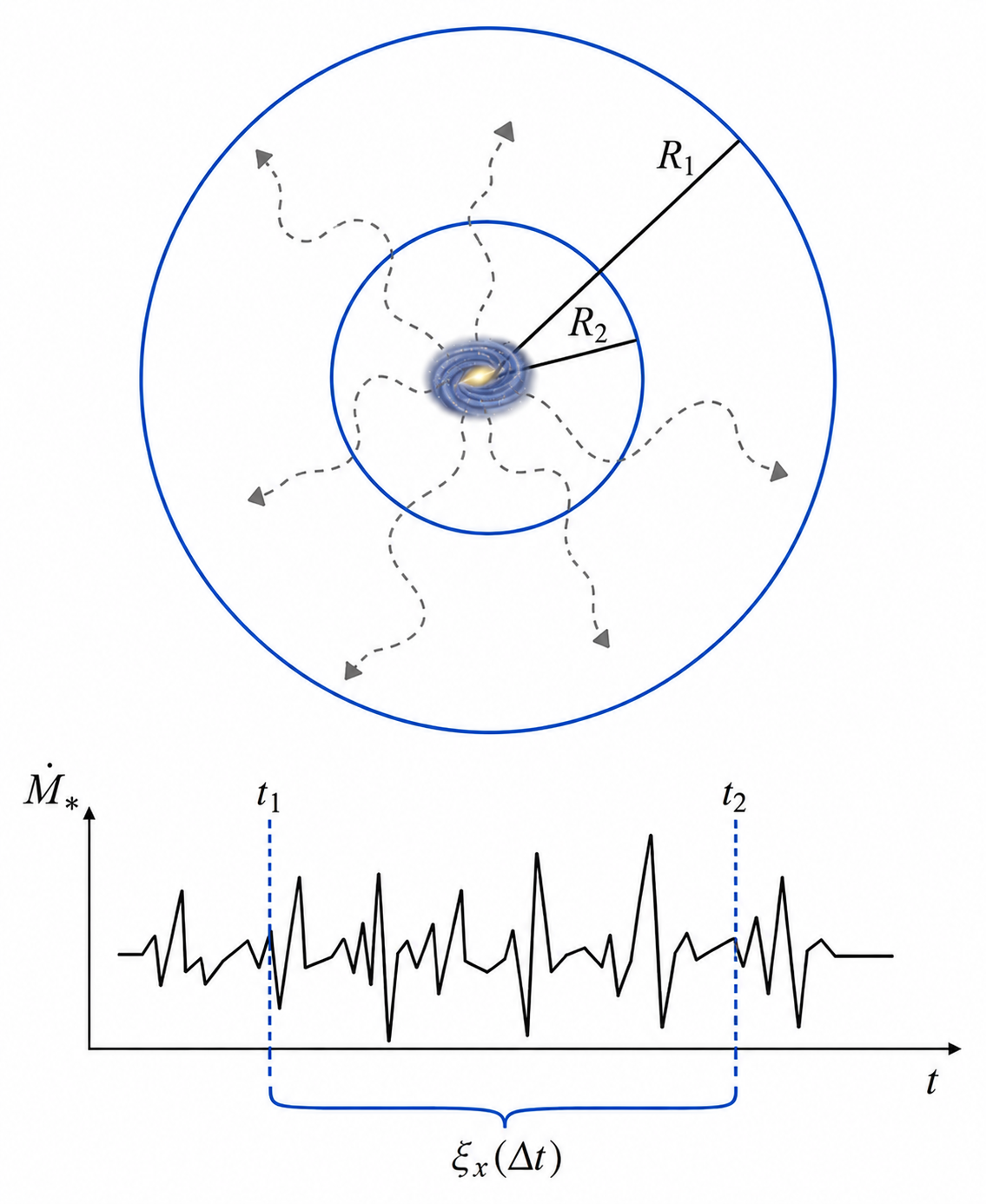}
    \caption{Illustration of the unequal-time correlation in our model. A central galaxy sources radiation fields according to the star formation history (SFH) shown at the bottom. A burst of star formation at time $t_1$ produces radiation that propagates outward; by the time a subsequent burst occurs at $t_2$, the radiation from the first burst has reached the shell at radius $R_1$. The correlation between the radiation fields at radii $R_1$ and $R_2$ depends on their separation relative to the coherence timescale of the bursts $\tau_{\rm PS}$, which determines the unequal-time correlation function $\xi_x(\Delta t)$    .}
    \label{fig:illustration}
\end{figure}

Figure~\ref{fig:illustration} illustrates this effect. The radiation that determines the 21-cm signal at a given position $\boldsymbol{x}$ is sourced by a series of concentric shells, each probing the SFRD at a different cosmic time. The contribution from any pair of shells therefore depends on the unequal-time SFRD correlation between the corresponding epochs, introducing a burst-induced shot-noise component analogous to Eq.~\eqref{eq:Pshot_master}. The resulting convolution is non-trivial and differs significantly between the Ly$\alpha$ and X-ray fields because of their distinct radiative kernels.

To compute these effects quantitatively, we adopt the \texttt{Zeus21}~\cite{Munoz:2023kkg} formalism with only PopII stars, although our results can be generalized to any framework based on Eq.~\eqref{eq:fluxes_general}. The standard implementation of \texttt{Zeus21} does not include shot-noise contributions arising from the discreteness of the halo population. We therefore first extend the formalism to incorporate the deterministic shot noise associated with discrete halos and subsequently include the additional burst-induced contribution arising from unequal-time SFR correlations. Appendix~\ref{sec:Zeus21} summarizes the standard \texttt{Zeus21} treatment, while the following Sections present these extensions and their impact on the predicted 21-cm signal.

\subsection{Lyman-$\alpha$ Contribution}\label{sec:LyA}

We build the local Ly$\alpha$ flux by integrating the SFRD contributions from spherical shells of radius $R$ around each point $\boldsymbol{x}$. Due to the light-travel time (assuming straight line motion, see Ref.~\cite{Flitter:2026emz} for a more advanced model of Ly$\alpha$ photon propagation), the flux received in $\boldsymbol{x}$ at $t(z)$ has been sourced from each shell at the retarded time 
\begin{equation}\label{eq:shell_time}
    t(z,R)\!=\!t(z)\!-\int_0^R \frac{d\chi}{c(1+\mathbf{z}(\chi))}
\end{equation} where $\chi$ is the comoving distance and $\mathbf{z}(\chi)$ the redshift associated with it; therefore, 
\begin{equation}
    J_\alpha(\boldsymbol{x}, z) \;=\; c_{\alpha,1}(z)\sum_R \Delta R\,\epsilon_\alpha(R)\,\dot{\rho}_*(\boldsymbol{x}, R, t(R))\Theta^R(\boldsymbol{x})\,,
\label{eq:zeus_lya_app}
\end{equation}
where $c_{\alpha,1}(z)$ collects normalization factors, while $\epsilon_\alpha$ is the Ly$\alpha$ emissivity (see Appendix~\ref{sec:Zeus21} for details). 

In the thin-shell approximation, the shell at $R$ is identified through $\Theta^{R}(\boldsymbol{x})$, which equals $1$ for $|\boldsymbol{x}|\!\in\![R, R\!+\!\Delta R]$ and $0$ otherwise; its angular
Fourier transform is
\begin{equation}
    \hat W_{R}(k) \;\equiv\; \int\! d^{3}\boldsymbol{x}\,\Theta^{R}(\boldsymbol{x})\,e^{-i\boldsymbol{k}\cdot\boldsymbol{x}} \;=\; 4\pi R^{2}\,j_{0}(kR),
\label{eq:Wshell_def}
\end{equation}
with $j_{0}(kR)=\frac{\sin(kR)}{kR}$ the zeroth-order spherical Bessel.

By taking the Fourier transform of the two-point function in Eq.~\eqref{eq:zeus_lya_app}, we obtain the following expression for the Ly$\alpha$ auto-power spectrum:
\begin{align}
P_{\rm Ly\alpha}(k,z) &\;=\; c_{1,\alpha}^{2}(z)\sum_{R_{1},R_{2}}\Delta R^{2}\,\epsilon_\alpha(R_{1})\,\epsilon_\alpha(R_{2})\nonumber\\
&\times\,\hat W_{R_{1}}(k)\,\hat W_{R_{2}}(k)\,\bigl\langle\hat{\dot{\rho}}_*(\boldsymbol{k}, t(R_{1}))\,\hat{\dot{\rho}}_*^\dagger(\boldsymbol{k}, t(R_{2}))\bigr\rangle\,.
\label{eq:Pjalpha_master_app}
\end{align}
Substituting the unequal-time SFRD cross-correlator characterized in Sec.~\ref{sec:shotnoise}, allows us to split the Ly$\alpha$ power spectum into
$P_{\rm Ly\alpha}(k,z)\!=\!P_{\rm Ly\alpha}^{\rm clust}(k,z)+P_{\rm Ly\alpha}^{\rm shot}(k,z)$, with the clustering term unchanged compared to the standard \texttt{Zeus21} formalism. On the other hand, for the shot noise
\begin{equation}
\begin{aligned}
P_{\rm Ly\alpha}^{\rm shot}(k, z) \;=\; c_{1,\alpha}^{2}(z) \sum_{R_{1}, R_{2}}\Delta R^{2}\,\epsilon_\alpha(R_{1})\,\epsilon_\alpha(R_{2})\,\\
\times \hat W_{R_{1}}(k)\,\hat W_{R_{2}}(k) P^{\rm shot}_{\dot{\rho}_*}(k; t(R_1), t(R_2)).
\label{eq:Pshot_lya_app}
\end{aligned}
\end{equation}
In the bursty kernel in the second line, similarly to Eq.~\eqref{eq:Pshot_master}, the mean SFR is evaluated at the two emission epochs separately, while the HMF is evaluated at the highest redshift.
The shell-convolution procedure introduces a $k$-dependent shot noise, which is maximized when $R_1=R_2$, i.e., equal time.

\subsection{X-ray Contribution}\label{sec:Xrays}

Computing the X-ray flux requires information on the past heating history: a single observation point at $z$ receives contributions from
many emission redshifts, because of the long X-ray mean free path. With respect to the Ly$\alpha$ case, this introduces an additional integral over redshift, $z'$. To add burstiness to the X-ray flux, we build the X-ray temperature contribution at $(\boldsymbol{x},z)$ 
from the SFRD at all earlier emission epochs $z'\!\geq\!z$,
\begin{equation}
\begin{aligned}
    T_X(\boldsymbol{x}, z) \;&=\; \sum_{z'\geq z}c_{1,X}(z')\sum_R \Delta R\,c_{2,X}(R,z'))\\
    &\times\dot{\rho}_*(\boldsymbol{x}, R, t'(z',R)),
\end{aligned}\label{eq:TX}
\end{equation}
where $c_{1,X}(z')$ collects the propagation and unit-conversion factors,
$c_{2,X}(R,z')$ is the X-ray radial coupling containing the X-ray emissivity, and $t'(z', R)\!=\!t(z')\!-\int_0^R {d\chi}/[{c(1+\mathbf{z}(\chi))}]$ represents the retarded time within emission
epoch $z'$ (note the difference with respect to Eq.~\eqref{eq:shell_time}). 
Similarly to what we discussed in Eq.~\eqref{eq:Pshot_master}, we identify the same physical halo at the two emission epochs by fixing its mass $M_h$ at the earlier redshift, $z_{\rm ref}\!=\!\max(z(t(z'_{1},R_1)), z(t(z'_{2},R_2)))$, and we forward-evolve it using the exponential accretion model.

Estimating the shot-noise power spectrum of $T_X$ from Eq.~\eqref{eq:TX}, therefore, requires evaluating a four-dimensional integral over pairs of emission redshifts $(z'_1,z'_2)$ and shell radii $(R_1,R_2)$. 
Summing over past redshifts and shells, we obtain that the X-ray power spectrum reads as 
\begin{align}\label{eq:Tx_4D_app}
P_{T_X}^{\rm shot}(z, k)&=\sum_{z'_{1}, z'_{2}\geq z}c_{1,X}(z'_{1})c_{1,X}(z'_{2})\sum_{R_{1}, R_{2}}\Delta R^{2}\nonumber\\
&\times c_{2,X}(R_{1}, z'_{1})\,c_{2,X}(R_{2}, z'_{2})\,\hat{W}_{R_{1}}(k)\,\hat W_{R_{2}}(k)\nonumber\\&\times P^{\rm shot}_{\dot{\rho}_*}(k; t(z_1',R_1), t(z_2',R_2)).
\end{align}
The previous equation contains the same shell-pair as
the Ly$\alpha$ kernel of Eq.~\eqref{eq:Pshot_lya_app}, but with two emission redshifts per shell-pair rather than one. In this way, the 4D-correlated kernel allows us to account for coherent contributions to the heating rate of the same halo at fixed comoving position but multiple emission epochs. 

\begin{figure*}[t]
\centering
\includegraphics[width=\textwidth]{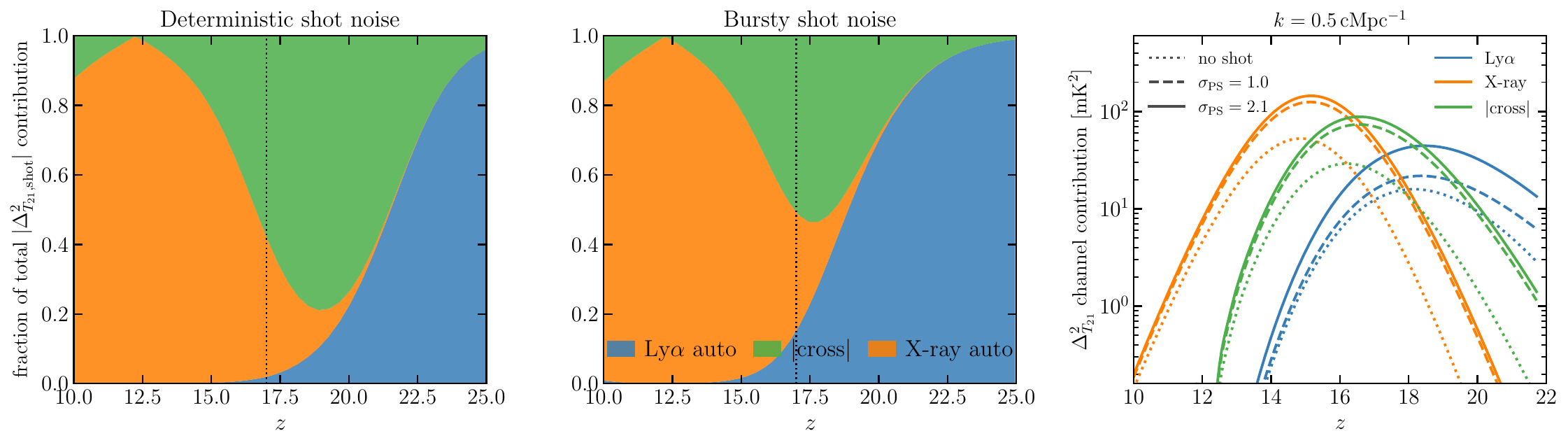}
\caption{{\it Left and Middle:} Stacked-area decomposition of Cosmic Dawn redshift evolution of $|\Delta^2_{T_{21},{\rm shot}}|$ at
$k = 0.5\Mpcinv$ into the three contributing channels: Ly$\alpha$
auto (blue), X-ray auto (orange), and the absolute value of the cross
(green). The left panel corresponds to the deterministic shot noise case ($\sigPS=0$) while the middle one presents the bursty case with the fiducial value from M26, $\sigPS=2.1$. 
{\it Right:} The contribution of each of the channels to the total $|\Delta^2_{T_{21}}|$, using the same color scheme as the left and middle panels, at $k=0.5 \rm \,cMpc^{-1}$. Similarly to the shot-noise component contribution, the total 21-cm power spectrum conserves the same time hierarchy, with Ly$\alpha$ dominating early, and X-rays taking over later on. In addition, the Ly$\alpha$ channel exhibits increased sensitivity to the burstiness amplitude, for two reasons. (i) The epoch where Ly$\alpha$ dominates is also characterized by stronger burstiness, as small burstier galaxies are more abundant. (ii) The Ly$\alpha$ radiation has a shorter traveling distance compared to X-rays, hence it is dominated by the contribution of the closest shells, for which $\Delta t \ll \tauPS$ and bursts are coherent. On the contrary, X-ray photons travel longer distances, hence distant shells receive radiation from very different epochs, for which the correlation between bursts is smoothed out.}
\label{fig:channel_sigma}
\end{figure*}

\subsection{Cross Contributions}
As detailed in Appendix~\ref{sec:Zeus21}, the \texttt{Zeus21} formalism includes contributions to the 21-cm power spectrum from the density field, as well as from the cross-correlations of the Ly$\alpha$ and X-ray fields with the density field and with each other. In our burstiness model, however, stochastic fluctuations in the star-formation rate are assumed to be independent of the large-scale environment. As a result, the burst-induced shot-noise contribution does not correlate with the density field, implying
\begin{equation}
P_{\delta}^{\rm shot}=0,\qquad
P_{{\rm Ly}\alpha,\delta}^{\rm shot}=0,\qquad
P_{T_X,\delta}^{\rm shot}=0.
\end{equation}
The only non-vanishing burst-induced shot-noise terms therefore arise from the Ly$\alpha$ and X-ray fields and their mutual cross-correlation, $P^{\rm shot}_{{\rm Ly\alpha},T_X}(k, z)$. Computing it involves a hybrid shell-pair integral with one window sourced by the bursty Ly$\alpha$ kernel at $R_1$ and one sourced by the X-ray kernel at $R_2$ at $z'>z$. Therefore, the cross term is given by
\begin{align}\label{eq:cross_3D_app}
P_{T_X,x_\alpha}^{\rm shot}(z, k)&=\sum_{z'\geq z}c_{1,\alpha}(z)c_{1,X}(z')\sum_{R_{1}, R_{2}}\Delta R^{2}\nonumber\\
&\times \epsilon_\alpha(R_1)\, c_{2,X}(R_{2}, z')\,\hat{W}_{R_{1}}(k)\,\hat W_{R_{2}}(k)\nonumber\\&\times P^{\rm shot}_{\dot{\rho}_*}(k; t(z,R_1), t(z',R_2)).
\end{align}
with $t$ defined as before.  

\subsection{Relative importance of Ly$\alpha$ and X-ray perturbations to the shot noise power spectrum}\label{sec:compare_shot}

To conclude this Section, we briefly comment on the relative importance of burstiness in the computation of shot noise in the Ly$\alpha$, X-ray and cross-power spectra. 

The bursty enhancement over the deterministic shot noise is dominated by the $|t(R_1) -   t(R_2)| < \tauPS$ contribution,  where the retarded times of the two shells is within the same burst; this is maximized when $R_1=R_2$, while off-diagonal pairs are suppressed by the decay of the OU correlator toward unity. In the case of Ly$\alpha$, whose photons have a mean free path of a few hundred Mpc,
a substantial fraction of the relevant shell pairs lies close to the diagonal, making the bursty enhancement a leading-order effect. By contrast, the longer mean free path of X-rays implies that most shell pairs lie far from the diagonal, where the OU kernel is essentially unity, so that the bursty enhancement constitutes only a small fractional correction on top of the deterministic shot noise. We note that the travel distance of the photons is encapsulated within $\epsilon_\alpha(R)$ and $\epsilon_X(R,z)e^{-\tau_X(z)}$ for Ly$\alpha$ and X-rays, respectively.

Despite their absolute amplitudes, Ly$\alpha$ and X-rays contribute to the final 21-cm power spectrum through the weighted sum in Eq.~\eqref{eq:power_21cm}; the amplitudes of the $\beta_i$ coefficients, which translate perturbations in the radiative fields into perturbations in $T_{21}$, therefore determine their relative importance in the total shot-noise budget. We discuss this in Fig.~\ref{fig:channel_sigma}, where the Ly$\alpha$, X-ray, and cross-contributions are weighted by the corresponding $\beta_i$ coefficients.

In our baseline case ($\sigma_{\rm PS}=2.1$, middle panel), the redshift evolution in the plot follows naturally from the physics of Cosmic Dawn:
\begin{itemize}
\item At {high} redshift ($z\!\gtrsim\!18$), the X-ray auto term is barely sourced, since before $z\simeq 20$ the cumulative heating background has not yet built up in the adopted model. Thus, the emission integral samples a small range of emission redshifts $z'\!\geq\!z$ and $\Delta^{2}_{T_X,\,{\rm shot}}$ is tiny; the Ly$\alpha$ channel and its cross with 
the (weak) X-ray field carry the signal.
\item At the WF-transition, when X-ray heating becomes relevant ($z\!\sim\!15$--$17$), the three
channels are comparable in magnitude, with X-rays leading at the lower-$z$ end of
this window and Ly$\alpha$ taking over at high $z$. The cross power spectrum is negative
throughout this window, so a substantial cancellation reduces the total below the sum of the
auto contributions.
\item At {low} redshift ($z\!\lesssim\!14$), the X-ray channel dominates. The
Ly$\alpha$ coupling has saturated, and its relative contribution to the final result
collapses even though $\Delta^{2}_{{\rm Ly}\alpha,\,{\rm shot}}$ remains large in its own
units. 
\end{itemize}

In the following Sections, we present the impact of the shot noise terms on the 21-cm power spectrum and discuss its implications and possible caveats. From the implementation point of view, we note that adding the bursty shot noise to the 21-cm power spectra slows down \texttt{Zeus21} by a factor of $\sim5$ with respect to the default version. The reason behind it is the 5D integral required to estimate the X-ray correlation, Eq.~\eqref{eq:Tx_4D_app}: in practice, this needs to account for pairs of emission redshifts $(z_1', z_2')$, shell radii $(R_1, R_2)$ and the halo mass ($M_h$) inside the SFRD power spectrum, see Eq.~\eqref{eq:Pshot_master}. Instead, the original code only requires a 4D integration (with the $M_h$ dimension removed), as shown in Appendix~\ref{sec:Zeus21}.

\begin{figure*}[t!]
\centering
\includegraphics[width=\textwidth]{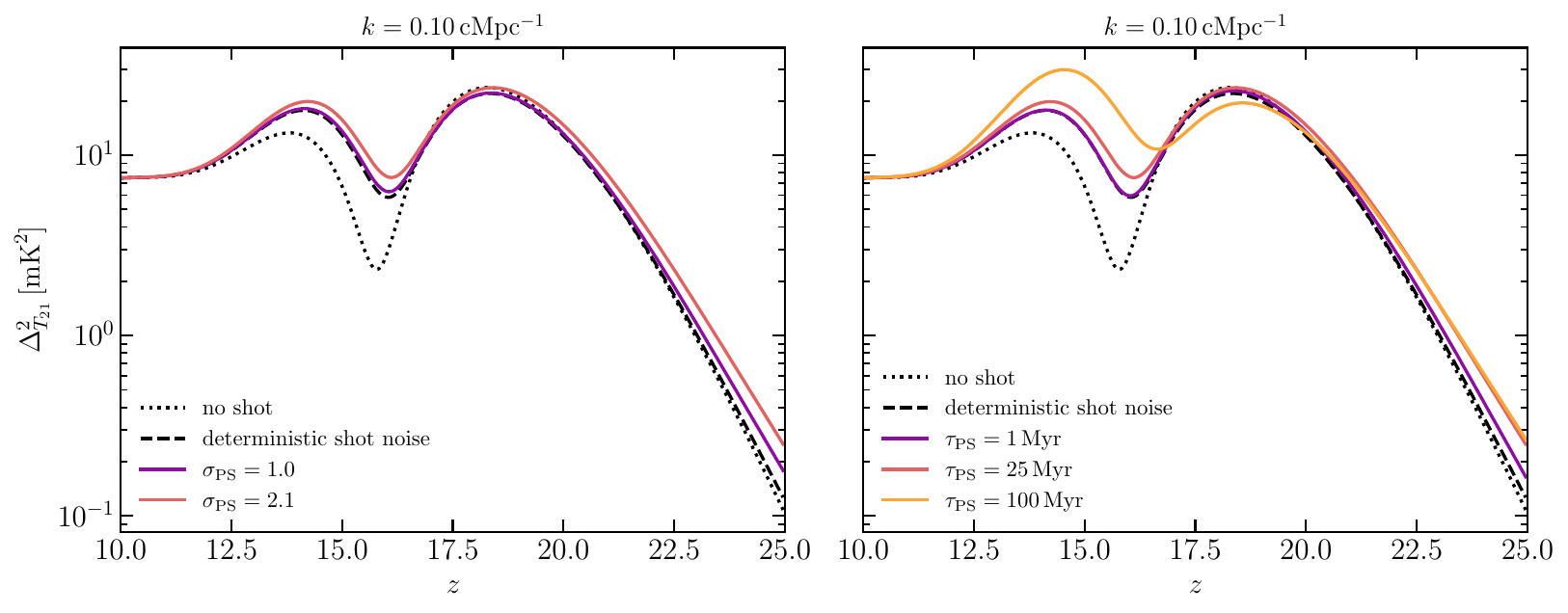}
\caption{{\it Left:} Total 21-cm power-spectrum as a function of redshift at
$k = 0.1\Mpcinv$, at fixed fiducial parameters, with $\sigPS=1.0 ,2.1$ which stand for moderate and fiducial burstiness respectively, and  {\it no-shot} (dotted) and {\it deterministic-shot-noise} (dashed) cases for reference. The shot noise introduces an enhancement at the Wouthuysen-Field (WF) dip even in the deterministic case, and is further increased with $\sigPS$. A smaller enhancement is evident at the two peaks, and the early Cosmic Dawn signal is boosted as well; we discuss the latter in detail in Sec.~\ref{sec:Discussion}. 
{\it Right:} Total 21-cm power-spectrum as a function of redshift at
$k = 0.1\Mpcinv$, at fixed fiducial parameters, where the  burst correlation time $\tauPS$ is varied. Increasing $\tauPS$ shifts the WF transition into earlier times, and significantly increases the amplitude of the signal at the end heating peak.
}
\label{fig:21cm_vs_sigma_tau}
\end{figure*}

\begin{figure*}[htbp]
\centering
\includegraphics[width=\textwidth]{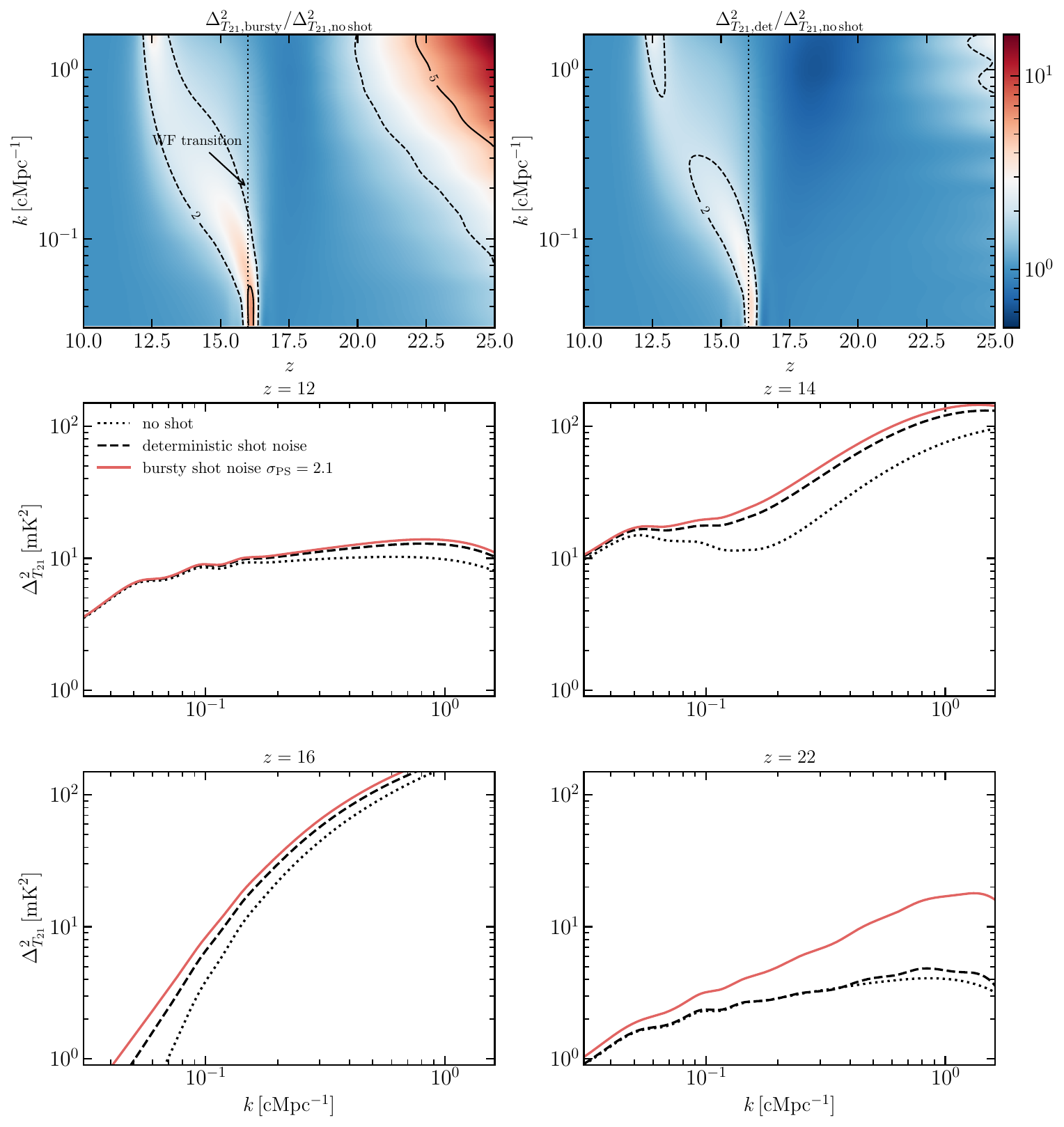}
\caption{
Impact of burstiness on the 21-cm power spectrum across Cosmic Dawn. The {\it four bottom panels} show $\Delta^2_{T_{21}}(k)$ at $z=\{12$, $14$,
  $16$, $22\}$, comparing three models: the {\it no-shot} baseline, the {\it deterministic-shot} noise floor set by the discreteness of halos when the average SFR is assumed, and the {\it bursty} model with $\sigma_{\rm PS}=2.1$. At $z=22$, well before the
  WF transition, the deterministic shot noise is sub-dominant on most scales, while burstiness boosts the signal by up to an order of magnitude on small
  scales, producing a shot-dominated tail that overwhelms the {\it no-shot} power. Fluctuations at such redshifts are dominated by halos whose burstiness is not directly constrained by the data used in M26. By $z=16$, during the WF transition, the boost is apparent on all scales, and even more pronounced at the large scale end where the transition dip is extremely sharp (see also at the bottom of the right top panel). At $z=14$ and $12$, well within the X-ray heating
  epoch, the three models converge on the largest scales (where clustering fluctuations dominate) and the bursty correction reduces to a small-scale boost. The {\it top row}
  presents the ratio maps in the $(z, k)$ plane: the {\it left panel}, $\Delta^2_{\rm bursty}/\Delta^2_{\rm no,shot}$, shows that bursty enhancement reaches
  factors of $2$-$5$ over a wide region of parameter space, peaking at high $z$ and small scales, where the clustering signal is weakest and small halos with strong burstiness dominate; the {\it right panel}, $\Delta^2_{\rm
  det}/\Delta^2_{\rm no,shot}$, shows that the deterministic shot noise alone produces a far smaller and more localized correction, only crossing a factor of $2$ in a narrow
  strip around the WF transition.
}
\label{fig:ratio_map}
\end{figure*}
\section{Results}\label{sec:results}

With all the ingredients in place, we now examine how the bursty shot noise reshapes the Cosmic Dawn 21-cm
power spectrum. We compare three physical configurations, from the bare clustering signal to the new full one: 
\begin{itemize}
\item[(i)] the {\it no-shot} baseline as implemented by the original \texttt{Zeus21}, in which the SFRD is treated as a smooth deterministic
field with no per-halo discreteness; 
\item[(ii)] the {\it deterministic-shot noise}, in which we treat halos as discrete sources, but each of them carries a smooth SFH, namely we consider the
$\sigPS \to 0$ limit. In this case, the 21-cm power spectrum receives a time-independent contribution from the Poisson shot noise; 
\item[(iii)] the {\it bursty-shot noise}, in which
the per-halo SFR is correlated over time trough the OU kernel, as described in Sec.~\ref{sec:shotnoise}. 
\end{itemize}
Unless otherwise stated, we assume the fiducial burstiness parameters  in Tab.~\ref{tab:fid}, cosmological parameters from Planck 2018~\cite{Planck:2018vyg} and the default \texttt{Zeus21} astrophysical parameters, fitted on UVLFs \cite{Sabti:2021unj, Sabti:2021xvh} and in agreement with the most recent EoR observations \cite{HERA:2021bsv, HERA:2022wmy}.

The left panel of Fig.~\ref{fig:21cm_vs_sigma_tau} shows the total (i.e.,~clustering + shot noise) 21-cm power spectrum 
$\Dttone(z)$, at a characteristic scale $k = 0.10\Mpcinv$. We observe that 
the Cosmic Dawn double-peak structure is preserved when adding {deterministic} and {bursty} shot noise: the heating
peak at $z \sim 14$ and the Ly$\alpha$ coupling peak at $z \sim 19$
bracket a sharp trough at the WF transition $\zWF \sim 16$. The
{\it deterministic-shot} curve lies very close to the {\it no-shot} baseline for most of the signal evolution, except inside the WF trough. Here, the signal gets boosted when shot noise is taken into account, as including a discrete-source noise term produces a noise floor that increases the minimal signal that can be measured at each redshift. 

The {\it bursty-shot} curve in the left panel of Fig.~\ref{fig:21cm_vs_sigma_tau} has a stronger impact than the deterministic one. While the two lines are close at the heating peak, they diverge toward the
WF transition, where the per-halo SFR autocorrelation induced by burstiness lifts the trough
by a factor of a few with respect to the baseline in the fiducial case. In addition, we obtain an increasing enhancement toward high $z$, where the signal is dominated by low-mass halos ($\sim10^9M_\odot$). Due to our modeling and given the burstiness amplitude slope calibrated by M26, such small halos exhibit increased burstiness and produce strong shot noise that overcomes the clustering signal at early times.
Finally, as the EoR approaches at $z\lesssim12.5$, the clustering signal takes over and the shot noise component becomes subdominant.
The ratio between the {\it bursty}, {\it deterministic-shot} and {\it no-shot} signals at all relevant scales and redshifts is presented in Fig.~\ref{fig:ratio_map}. 

The right panel of Fig.~\ref{fig:21cm_vs_sigma_tau} isolates the effect of the star formation correlation time on the evolution of the 21-cm power spectrum. Varying $\tau_{\rm PS}$
controls the radial range over which the bursty
amplification operates (see Eq.~\eqref{eq:Pshot_master}); in the short burst limit ($\tauPS = 1\,{\rm Myr}$), the curve is
nearly indistinguishable from the baseline as the kernel
decorrelates faster than most of the shell-pair light-travel times and the OU enhancement averages away. At
$\tauPS = 25\,{\rm Myr}$, the fiducial value we inherit from M26, the bursty boost emerges
across the full redshift range, with the WF trough lifted by a factor of the order of a few. In the limit of very long bursts ($\tauPS = 100\,{\rm Myr}$), the heating peak grows by
roughly a factor of two relative to the deterministic curve, and notably the WF transition is shifted toward earlier times due to the enhancement in the X-ray field, introduced by the OU autocorrelation kernel  (see Eq.~\eqref{eq:y_autocorr}), which now includes a significant amount of correlated shells.

The full scale and redshift dependence of the bursty enhancement is
summarized in Fig.~\ref{fig:ratio_map}, which shows the ratio
$\Delta^2_{T_{21},\rm bursty}/\Delta^2_{T_{21},\rm  no\,shot}$ and $\Delta^2_{T_{21},\rm det}/\Delta^2_{T_{21},\rm  no\,shot}$ 
as a function of $(z, k)$ (top panels), and vertical slices of those panels at characteristic redshifts,
at the M26 fiducial parameters. Three regions of the $(z, k)$ plane
stand out. First, the ratio is bounded near unity throughout the
post-WF coupling regime ($z \lesssim \zWF$), where the clustering
signal dominates and the bursty shot-noise component contributes at the few-percent level. 
Second, a feature at $z \sim 16$ that marks the WF transition is apparent on all scales. The difference in the ratio of the signals between the small and large scales, mostly stems from the depth of the WF trough, which widens on small scales while extending toward lower redshifts. 
Third, the ratio rises steeply with
redshift in the early Cosmic Dawn for small scales: at $k \gtrsim 0.3\Mpcinv$
and $z \gtrsim 22$ it exceeds 5, and reaches an order of magnitude
near $z \sim 25$. This is the regime where the underlying signal is
small, the source population is dominated by small mass halos and the mass-dependent OU amplitude $\sigPS$ is saturated at its maximal value, providing a significant amplification. At larger $k$s, the contribution of these small halos is smoothed out and the ratio of the bursty and the no-shot signal does not exceed $\sim 3$. 
We emphasize that the scale dependence of the shot noise enhancement is determined by the extent to which the signal can trace the discreteness and stochasticity of the underlying sources. On large scales, these fluctuations are averaged over many halos, and the impact of any individual source is therefore suppressed. On smaller scales, however, where the probed $k$s become comparable to the typical separation between sources, shot noise becomes increasingly important and can substantially enhance the signal.
Finally, contrasting the top-left and top-right panels of Fig.~\ref{fig:ratio_map} allows us to assess the relative importance of shot noise in the {\it bursty} and {\it deterministic} cases. 

Comparing the result we obtained here for the 21-cm signal to the one obtained for other LIM observations (H$\alpha$, H$\beta$, CO, [CII],...) in Ref.~\cite{Kovetz:2026xfs}, we find that the 
enhancement induced by burstiness in $\Delta^2_{T_{21}}$ is less pronounced than the boost observed on small scales in the LIM power spectra. This difference arises because the 21-cm signal depends on radiation fields that average over extended radial shells. As a result, the discreteness of the underlying halo population is partially smoothed out. By contrast, the shot-noise contribution to LIM observables is intrinsically local, making it more sensitive to the stochasticity of individual sources. Consequently, the burst-induced enhancement is partially washed out in the 21-cm power spectrum relative to the LIM case.

\section{Discussion}\label{sec:Discussion}

\begin{figure*}[t]
\centering
\includegraphics[width=\textwidth]{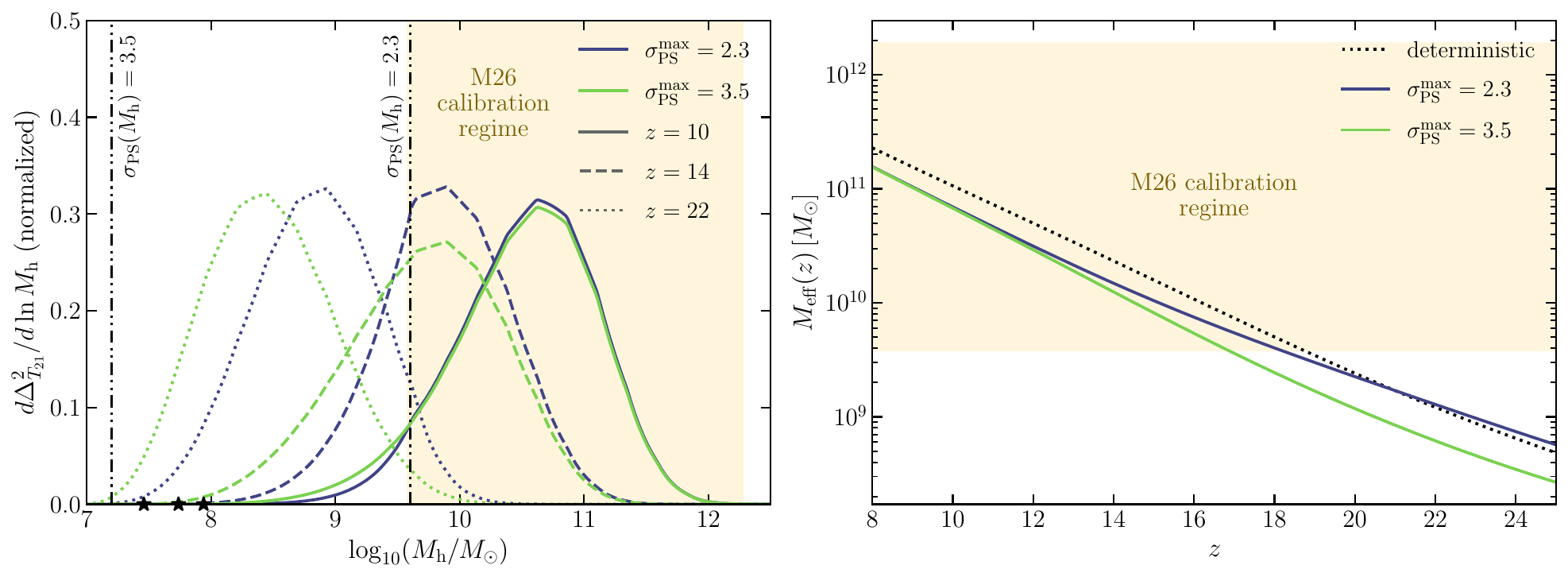}
\caption{{\it Left}: Halo-mass dependence of the bursty 21-cm shot-noise integrand, $d\Delta^2_{T_{21}}/d\ln M_{\rm h}$ (normalized), at different redshifts
  for two caps: $\sigPS^{\rm max} =2.3$, which corresponds to the value reached at the smallest halos probed by M26 data, and $\sigPS^{\rm max} =3.5$ which allows the amplitude to grow up to the smallest halos probed by 21-cm data (considering only PopII). 
  The vertical markers
indicate where $\sigPS(\Mh)$ crosses each cap if extrapolated without truncation; below
those masses the OU factor is fixed at $e^{(\sigmax/\sqrt2)^2}$ for each of the caps. The three markers on the $x$-axis at $M_h\lesssim 10^8M_\odot$ correspond to the atomic cooling threshold~\cite{Munoz:2021psm}, below which galaxy formation is exponentially suppressed  for the redshifts considered here. As the threshold grows with $z$, the rightmost marker corresponds to $z=10$ and the leftmost to $z=22$.
{\it Right}: The halo mass that governs the integrand as a function of redshifts for different $\sigmax$ values, with the {\it deterministic-shot} case for reference. 
  For low redshifts, the value of the $\sigmax$ has a minimal effect on the kernel, while for higher $z$ the kernel shifts toward lower $M_h$ by a factor of a few.
  For both figures the regime of halo masses probed by the luminosity functions data and $\rm H\alpha/UV$ PDFs used by M26 to constrain the burstiness parameters, is indicated by a shaded yellow region.}
\label{fig:kernel_Mh}
\end{figure*}

\begin{figure}[t]
\centering
\includegraphics[width=\columnwidth]{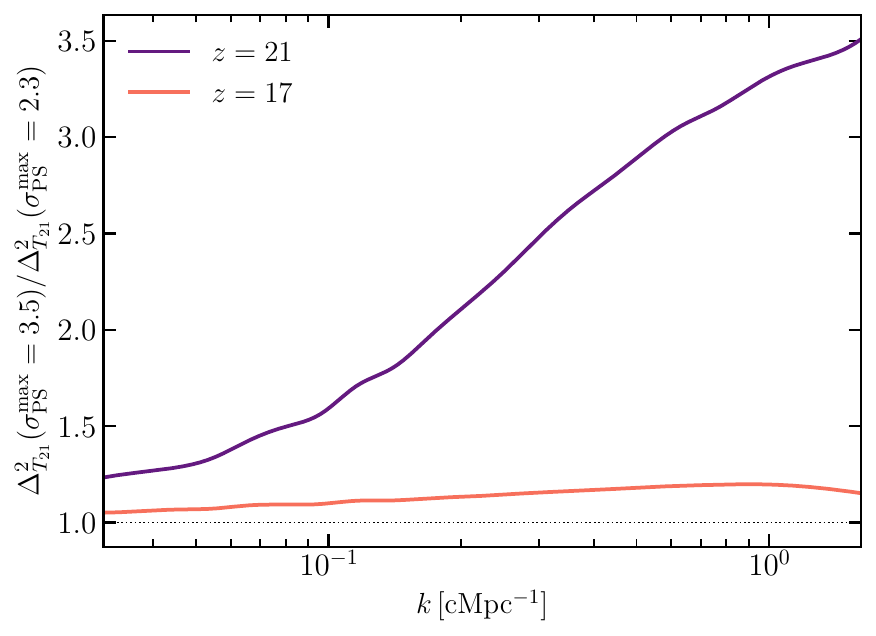}
\caption{The ratio of the total 21-cm power-spectrum with a cap on the burstiness amplitude at $\sigmax=\{2.3,3.5\}$ as a function of wavenumber at
$z=\{17,21\}$, at fixed fiducial parameters. Increasing $\sigmax$ mostly affects the amplitude of the signal at early times by allowing the low-mass tail of the OU amplification to grow, boosting it by up to a factor of $\sim 3.5$ at small scales. At later times, large halos dominate the Ly$\alpha$ and X-ray budgets, hence this impact is less pronounced.}  
\label{fig:sigmamax}
\end{figure}

Our results in Sec.~\ref{sec:results} show that burstiness contributes additively to the 21-cm power spectrum, on top of the deterministic clustering and shot-noise terms. The amplitude of this enhancement depends on the convolution between the radial-window support of the Ly$\alpha$ and X-ray fluxes, which defines their shell integrals, and the strength and mass dependence of burstiness, encoded in $\sigma_{\rm PS}(M_h)$. The interplay between these quantities determines both the amplitude and redshift evolution of the burst-induced shot noise.

As illustrated in Eq.~\eqref{eq:sigPS_M} and Fig.~\ref{fig:sigma_Mh}, the mass dependence of $\sigma_{\rm PS}(M_h)$ is characterized by the two parameters $d\sigma_{\rm PS}/d\log_{10}M_h$ and $\sigma_{\rm PS}^{\rm max}$. The slope is tightly constrained by M26, largely thanks to the H$\alpha$/UV measurements, whose increasing scatter toward the faint end favors stronger burstiness in lower-mass halos. Their analysis yields $d\sigma_{\rm PS}/d\log_{10}M_h=-0.5\pm0.13$. To assess the robustness of our conclusions, we additionally explore the extreme cases $d\sigma_{\rm PS}/d\log_{10}M_h=\{-1,0\}$. We find that given our conservative cap $\sigmax=2.3$, taking a larger slope $d\sigma_{\rm PS}/d\log_{10}M_h=-1$ hardly changes the signal at the relevant scales across Cosmic Dawn. This is because most of the halo masses below the pivot mass ($10^{10} M_\odot$) are saturated to the cap value, so the slope through which they approach the cap does not affect the signal. Taking the limit where the amplitude does not depend on the halo mass, $d\sigma_{\rm PS}/d\log_{10}M_h=0$, introduces a small suppression of the shot noise contribution, mostly at the high-$z$ end, dominated by small halos, which in this case would not feature enhanced burstiness. 
    
Varying astrophysical parameters can also alter the importance of the burstiness component. Our baseline assume parameters calibrated on the UVLF; we checked that increasing the slope of the 
low-mass end 
of the star formation efficiency (see Appendix~\ref{sec:Zeus21}), decreases the SFR in small halos, diminishing the impact of the strong burstiness our model associates with them. In contrast, flattening the slope increases the SFR in small halos, making the impact of their burstiness more pronounced.

\subsection{Mass Kernel and $\sigma_{\rm PS}^{\rm max}$}

The main limitation of propagating the M26 framework into 21-cm cosmology is the halo-mass range probed by the observational data used in the fit. As discussed in Sec.~\ref{sec:bursty}, these observations constrain burstiness only for halos with $M_h\gtrsim 5\times10^9\,M_\odot$. Because the preferred trend is an increased burstiness for lower masses, extrapolating the model beyond this range can lead to unrealistically large values of $\sigma_{\rm PS}$. To mitigate this issue, we introduced the $\sigma_{\rm PS}^{\rm max}$ cap, which effectively parameterizes our uncertainty in the burstiness of faint galaxies. Such a cap is physically motivated by the finite gas reservoirs of low-mass halos: while bursty episodes can temporarily elevate the SFR, sustaining arbitrarily large fluctuations would require converting gas into stars faster than it can be replenished through accretion~\cite{Bouche:2009gm,Dave:2012}. Extrapolating $\sigma_{\rm PS}$ to arbitrarily large values would thus imply fluctuations difficult to reconcile with the available gas budget.

Since current observations provide little guidance on the value of $\sigma_{\rm PS}^{\rm max}$, in our baseline analysis in the previous Section we saturate $\sigma_{\rm PS}$ right beyond the M26 boundary, adopting the conservative value $\sigma_{\rm PS}^{\rm max}=\sigma_{\rm PS}(M_h\approx5\times 10^9\,M_\odot)=2.3$. However, a different choice  may affect the predicted 21-cm signal, depending on which DM halo masses dominate the burst-induced shot noise.
To identify them and understand how well their burstiness is constrained by the data, it is useful
to look at the per-ln-mass integrand of the shot noise (see also Eq.~\eqref{eq:Pshot_master}). After
factoring out the redshift-independent geometric and radiative weights,
the contribution from a logarithmic halo-mass bin from the OU kernel scales as
\begin{equation}\label{eq:W_kernel}
    \mathcal{W}(\Mh, z) \propto \frac{dn}{d\ln \Mh}\,
                    \overline{\dot{M}}_*^2(M_h)
                    \exp\!\bigl[\sigxsq(\Mh)\bigr],
\end{equation}
where we set the SFR and OU kernels to equal time in order to isolate the dependence on the halo-mass.
Fig.~\ref{fig:kernel_Mh} shows $\mathcal{W}(\Mh, z)$ at three representative
redshifts for different $\sigmax$ values, and the halo mass that dominates the kernel, i.e.,
$M_{\rm eff}(z) = \int \mathcal{W}\,\Mh\,d\ln \Mh / \int \mathcal{W}\,d\ln \Mh$.
While the HMF favors small halos, the quantities defining the SFR enhance or suppress the importance of different mass ranges. The star formation efficiency (see Eq.~\eqref{eq:SHMR}) increases the importance of large halos below the peak mass parameter $M_p=3\times 10^{11} M_\odot$. In addition, the duty cycle $f_{\rm duty}$ enforces a redshift-dependent cutoff mass to account for inefficient star formation on the very low-mass end.
Finally, the OU kernel increases the impact of small halos through the halo mass slope, and can reach extreme values if not capped at $\sigmax$.

In the left panel of Fig.~\ref{fig:kernel_Mh}, we present the $\mathcal{W}$ kernel at different redshifts in the Cosmic Dawn to demonstrate how the shot noise depends on the cap value $\sigmax$. Its effect is particularly relevant for low mass halos, below the M26-constrained region (yellow area), where the exponential growth of the OU kernel alters the shape of the $\mathcal{W}$ kernel. 
At very high redshifts ($z\sim20$), these halos govern the 21-cm evolution, hence the choice of $\sigmax$ is crucial, as indicated in Fig.~\ref{fig:sigmamax}.  The right panel of Fig.~\ref{fig:kernel_Mh} shows that the regime of halo masses dominating the burstiness kernel can shift by a factor of a few at early times if the cap is increased from $2.3$ to $3.5$. This is of great importance for the shot noise signal, as shown in Fig.~\ref{fig:sigmamax}: enhanced burstiness in small halos significantly increases the 21-cm power spectrum at $z\sim 21$. At later times, around the WF transition and thereafter, increasing the cap has a smaller impact, as the halos that exhibit boosted burstiness have a subdominant contribution to the signal.
Increasing the cap to a large value such as 3.5, allows for smaller values of the slope $d\sigma_{\rm PS}/d\log_{10}M_h$ (e.g. $-1$) to affect the signal, and increases it in the high redshift regime. 

The situation changes at low redshifts ($z\lesssim z_{\rm WF}$), where the shot noise is dominated by halos big enough to be probed through current data; here, constraints from M26 can be safely forward propagated into the 21-cm power spectrum. At such redshifts, $\sigmax$ has no strong implications on the $\mathcal{W}$ kernel, controlled by halos large enough to not be affected by the cap. This confirms
that the bursty enhancement reported in
Figs.~\ref{fig:21cm_vs_sigma_tau}\nobreakdash--\ref{fig:ratio_map} for $z \lesssim z_{\rm WF}$ is
sourced by constrained halos, and is not
an artifact of an unbound extrapolation. 
Current 21-cm experiments, such as HERA~\cite{DeBoer:2016tnn, HERA:2021bsv, HERA:2022wmy} and LOFAR~\cite{LOFAR:2013jil, Mertens:2020llj, Ceccotti:2025bcd}, indeed target this redshift range, where the bursty shot noise signal can be safely determined. Future experiments such as SKAO~\cite{Mellema:2012ht, 2010arXiv1008.2871G}, will probe higher redshift; thus, it will be important to set physical bounds on $\sigmax$ in order to obtain the effect of burstiness on the signal without the risk of overestimating it.

\subsection{Sensitivity to $\dtau$}\label{sec:tau_evol}

\begin{figure*}[t]
\centering
\includegraphics[width=\textwidth]{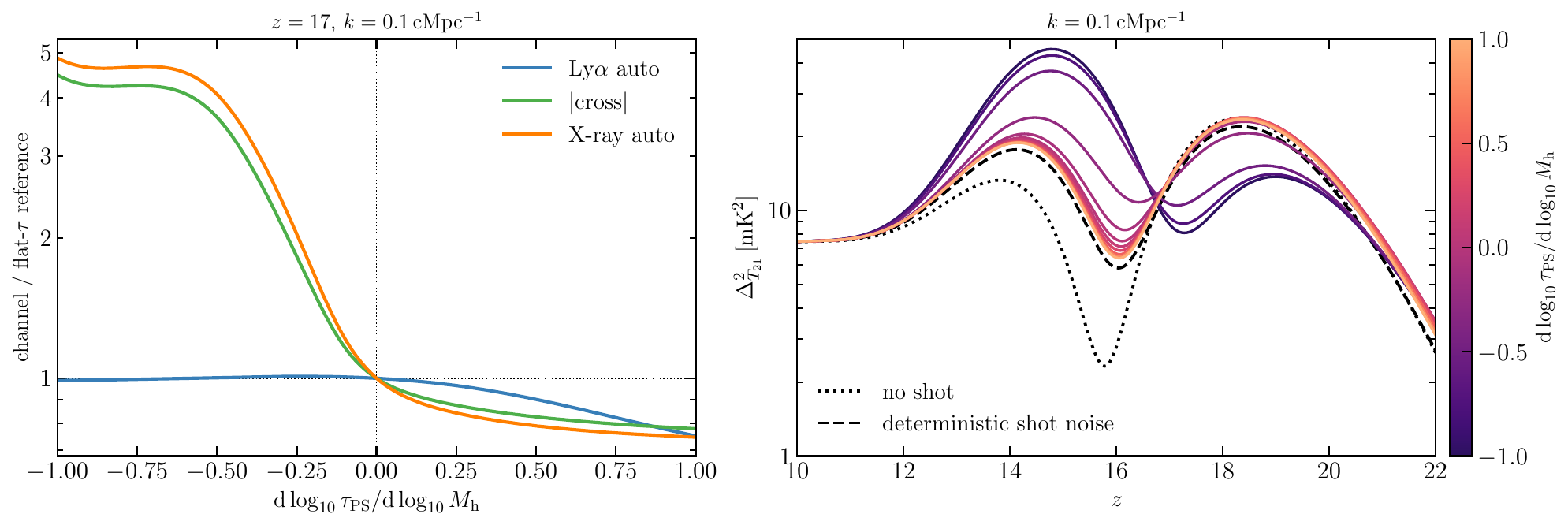}
\caption{{\it Left}: The impact of varying the halo-mass slope of the coherence time, $\dtau$, on each of the shot-noise channels in the Cosmic Dawn 21-cm power spectrum. The Ly$\alpha$ shot-noise power spectrum is almost insensitive to the difference in the slope, as it corresponds to nearby sources, with travel time smaller than the burst coherence time. The X-ray and the cross channels, instead, are affected by sources that are further apart from each other, making their shot noise highly dependent on the burst coherence time.
{\it Right}: The total 21-cm power spectrum at $k=0.1 \rm \, cMpc^{-1}$ for different values of $\dtau$. The bursty contribution to the signal is erased when the slope approaches 1, and mostly the {\it deterministic-shot} component survives, as burstier, small halos decorrelate faster. On the other hand, the X-ray heating epoch occurs earlier when $\dtau$ approaches $-1$, thanks to the longer correlation time for small halos with large $\sigPS$. This leads to an enhanced contribution from distant sources, amplifying the signal amplitude at the heating peak.}
\label{fig:dtau_sensitivity}
\end{figure*}
As mentioned in Sec.~\ref{sec:bursty}, M26 constrains the burst coherence time $\tauPS$ together with its halo-mass slope; however, the data they rely on carry little information on how the correlation time varies as a function of $M_{\rm h}$, $\dtau$. 
We now test the sensitivity of the Cosmic Dawn 21-cm power spectrum to this parameter, in order to understand the prospects for constraining it for a future joint inference. 
  
Figure~\ref{fig:dtau_sensitivity} isolates the effect of $\dtau$, fixing all other parameters. The left panel, evaluated at $z=17$ and $k=0.1\,\mathrm{cMpc^{-1}}$, shows the
shot-noise contribution of each coupling channel relative to the flat-$\tauPS$ case ($\dtau\!=\!0$). The behavior is strongly channel dependent; the Ly$\alpha$ auto-term
is nearly independent of $\dtau$, changing by  $\sim 20\%$  for extreme positive slopes (i.e.,~shorter coherence time for small halos), whereas the X-ray auto-term and the cross-term change by up to a factor of $\sim5$,
rising above unity for $\dtau<0$ and falling to $\sim0.5$ for $\dtau>0$.

This split follows directly from the light-travel-time structure of the two channels.
Ly$\alpha$ coupling is sourced by a relatively thin shell set by the
Lyman-series frequency band, so the two correlated bursts are separated by a
light-travel time $\Delta t \ll \tau_{\rm PS}$. In this limit the burst kernel
$\exp[\sigma_{\rm PS}^2\exp(-\Delta t/\tau_{\rm PS})]$ saturates at its maximum
$\exp(\sigma_{\rm PS}^2/2)$ and becomes insensitive to the value of $\tau_{\rm PS}$, and
hence to its mass dependence. X-ray heating, by contrast, is cumulative over a broad range of source redshifts, so the relevant $\Delta t$ is comparable to or larger than
$\tau_{\rm PS}$. In this regime, the kernel depends steeply on $\tau_{\rm PS}$: a negative
slope lengthens the correlation time of the low-mass halos that in our fiducial model exhibit stronger bursts, and therefore 
dominate the shot-noise term, increasing the correlation over the long X-ray baseline and
enhancing the power. On the other hand, a positive slope leads to a faster decorrelation in small halos, suppressing the shot-noise signal component. The
cross term inherits the same X-ray-driven sensitivity.

The right panel of Fig.~\ref{fig:dtau_sensitivity} translates this into the total 21-cm power spectrum. We find that varying this
unconstrained parameter reshapes the signal significantly.
For $\dtau\to1$, the signal collapses toward the {\it deterministic shot-noise} reference, since the contribution of each of the channels is diminished.
For $\dtau\!\to\!-1$, the power spectrum features are shifted toward earlier times, and enhanced strongly, as X-rays begin to impact the 21-cm power spectrum much earlier. A similar effect occurs when $\tauPS$ is flat but  increased,~e.g.,~$\tauPS=100 \rm \, Myr$ in the right panel of Fig.~\ref{fig:21cm_vs_sigma_tau}. 

We conclude that the halo-mass dependence of the burst timescale, which the UVLF and $\rm H\alpha/UV$ data used by M26
leave unconstrained, imprints a large and characteristic signature on the Cosmic Dawn 21-cm
power spectrum. The signal thus provides complementary leverage on the physics of
bursty star formation; data from future experiments will be used together with JWST observations to bound the model.

\subsection{Comparison to current data and detection prospects}
The bursty enhancement does not change the experimental landscape at $z\!\lesssim\!10$,
where current interferometers have set upper limits on $\Delta_{21}^2$~\cite{HERA:2022wmy,Lazare:2023jkg,Ghara:2025xzu}. At
$z\!\simeq\!10$ and $k\!\simeq\!0.3$\,-\,$1.0\,{\rm cMpc}^{-1}$, our predicted bursty power spectrum is $\Delta_{21}^2\!\sim\!10$--$20\,{\rm mK}^{2}$ (in the fiducial astrophysics); this value is consistent with the
current HERA Phase I upper limits and initial results from HERA Phase II,  
and provides no immediate observational tension. At the same time, our implementation of burstiness modifies only the fluctuations and leaves the global signal unchanged, and therefore our scenario does not alter the interpretation of sky-averaged measurements such as those from EDGES~\cite{Bowman:2018yin} and SARAS~\cite{Singh:2021mxo}.

Beyond the current limits, the interesting regime
for distinguishing bursty from non-bursty predictions via the 21-cm signal is the WF-transition epoch in the power spectrum
($z\!\sim\!15$\,-\,$16$). Here, the absolute $\Delta_{21}^2$ at
$k\!=\!0.1\,$cMpc$^{-1}$ grows from $\!\sim\!2\,{\rm mK}^{2}$ to $\!\sim\!9\,{\rm mK}^{2}$ between the {\it no-shot} and the {\it bursty} regimes. To detect such a difference, a substantial sensitivity improvement over HERA Phase I is required; The full HERA Phase II~\cite{2024PASP,HERA:2025ajm} and SKA-Low Phase 1~\cite{2010arXiv1008.2871G} are expected to reach this sensitivity regime at least at low Cosmic Dawn redshifts ($z\!\sim\!10-13$), depending on foregrounds assumptions, while 
extending toward $z\!\sim\!15$ with degraded sensitivity. 

Finally, the high-redshift regime, which is driven by low-mass halos in the $\sigma_{\rm PS}^{\rm max}$-sensitive range discussed in previous sections, lies beyond the reach of current interferometers. Because the 21-cm signal provides one of the few probes of star-formation activity in such faint and otherwise inaccessible galaxies, these results further motivate next-generation Cosmic Dawn experiments.

\subsection{Comparison with previous results}\label{sec:21cmFast_comparison}

In the latest version of the semi-numerical state-of-the art code $\texttt{21cmFAST}$~\cite{Davies:2025wsa} (hereafter D25), which simulates the 21-cm signal from the Cosmic Dawn to the EoR, a stochastic halo sampler was introduced, effectively incorporating a shot-noise component into the simulation. Specifically, D25 populate the simulation grid with halos, associating them with stochastic astrophysical properties (SFR, stellar mass, and X-ray luminosity) drawn from lognormal distributions. In addition, these attributes are correlated in time between simulation time steps, introducing an effect that resembles burstiness.

In the analysis presented in D25, the authors restrict themselves to a relatively small scatter, $\sigma_x\lesssim 0.5 \,\rm dex$, while in our case, inspired by the M26 constraints, we consider scatter up to $\sim 1 \, \rm dex$ for halos $\lesssim 10^{10} M_\odot$. Figure~9 in D25 presents the 21-cm power spectrum as a function of $k$ at two distinct epochs:  the mid point of the EoR, where the neutral fraction is $\overline{x}_{HI} = 0.5$, and at the X-ray heating epoch, when the global signal transitions from absorption to emission $(T_{21} \approx 0)$. While the former is out of reach for our work, the latter can be compared with the results we obtained in our analytic computation.  
Comparing the orange curve (no halo sampler) with the green (with halo sampler, no scatter in halo properties, i.e. similar to the model dubbed here as {\it deterministic}), and the red curves (with halo sampler and stochastic properties, with mean normalization, similar to our {\it bursty} model) enable us to examine whether the same effects we report here are also present in D25. 

On small scales, indeed a shot-noise-like effect boosts the signal above the case where the grid is not populated with halos, similarly to our results. On the other hand, contrary to our work, the {\it deterministic} and the {\it bursty} models seem to agree on all scales, and cannot be distinguished by eye. D25 attributes this phenomenon to the fact that the scattering in halo properties is small compared to the stochasticity in the number and mass of halos, introduced by the halo sampler, and is thus negligible. Taking a larger scatter by a factor of 2-3 from D25 fiducial values will produce a clear enhancement of the signal, which diverges from the {\it deterministic} model, as shown in Fig.~10 of D25.
In our case, where we do not populate a grid with halos, and the fiducial scattering is much larger than the fiducial D25 model, there is a prominent difference between the {\it deterministic} and {\it bursty} curves, as shown in  Fig.~\ref{fig:ratio_map}. 

On large scales, however, the curves in D25 that include stochastic halos show a mildly weaker power spectrum than the grid-based approach. In principle, and as we show here, adding a single-halo contribution should 
increase the fluctuations in both the Ly$\alpha$ and X-ray fields, where it enters as a purely additive shot-noise term. This enhancement should then propagate to the 21-cm power spectrum. While the latter receives contributions from several radiation fields with different weights, and their cross-correlations can in principle either enhance or suppress the total signal, these cross terms cannot compensate for the additional auto-power introduced by the single-halo contribution. We therefore expect the overall 21-cm power spectrum to increase rather than decrease. However, since the halo-sampler method in \texttt{21cmFAST} is completely different from defining a conditional HMF on the grid, variations are expected.

Another useful comparison is with Ref.~\cite{Reis:2021sqh}, which studied the impact of Poisson shot noise (analogous to our {\it deterministic-shot} case) and scatter in astrophysical parameters on the 21-cm power spectrum. The latter is qualitatively related to our burstiness scenario, although in their case the stochasticity arises from a random scatter in the astrophysical parameters rather than from time-dependent bursts. This can be interpreted as analogous to our model in the limit where the burst coherence time becomes infinite, so that each halo is assigned a constant, randomly drawn offset over its entire history. Their Fig.~5 shows the 21-cm power spectrum with and without shot noise, and both the WF trough feature and the high-redshift enhancement discussed in this work are clearly visible. Using simulations, they further demonstrate that introducing scatter in the astrophysical parameters increases the variance of the resulting 21-cm signal across realizations. This effect is not captured in our analysis, since we adopt an analytic framework rather than performing a suite of simulations. Nevertheless, their results show that increasing the stochasticity of the astrophysical parameters produces a larger impact than the deterministic shot-noise contribution alone, in qualitative agreement with the behavior found in our burstiness model.

\subsection{Further extensions to our analysis}\label{sec:extensions}

Throughout this work, we adopted some assumptions that can in principle be lifted to generalize the analysis. While leaving a detailed implementation for future work, we comment on them in this Section. 

First, our calculation assumes the M26 OU process is
a complete description of SFR stochasticity. 
In reality, star formation exhibits variability on multiple timescales, and incorporating these into our formalism would in part modify the results presented above. In particular, the UV-bright phase responsible for Ly$\alpha$ emission lasts only $\sim\,10\,\mathrm{Myr}$ after a burst, reflecting the short lifetimes of the massive stars that dominate the emission. Consequently, the Ly$\alpha$ radiation shells in our formalism would only be sourced over this limited time interval, reducing the temporal correlation between shells separated by large distances. This is analogous to the effect of finite quasar lifetimes discussed in Ref.~\cite{Meiksin:2018wij}, where a time-dependent radiative transfer formalism shows that accounting for the finite source lifetime suppresses the shot-noise contribution on large scales (low $k$). In the M26 framework, finite stellar lifetimes are already incorporated through Green's functions calibrated on stellar population synthesis models. A similar treatment could be applied to the Ly$\alpha$ emissivity, $\epsilon_\alpha$, and, in analogy to Ref.~\cite{Meiksin:2018wij}, would likely reduce the large-scale burstiness signal, bringing it closer to the {\it deterministic-shot} prediction. Our main results, however, concern small scales, where nearby radiation shells dominate the signal; we therefore expect our qualitative conclusions to remain unchanged. Meanwhile, the X-ray contribution is more involved. Since we assume X-rays are produced by stellar remnants, there is an intrinsic delay between a star-formation burst and the onset of X-ray emission. Moreover, the X-ray emission persists over much longer timescales, giving rise to a broad temporal kernel that may overlap contributions from multiple bursts. Modeling these effects consistently requires extending the present formalism, which we leave to future work.

Second, as previously anticipated, we have assumed per-halo independent
stochasticity. If bursts are environmentally triggered, e.g.,~by mergers, filamentary
accretion, large-scale assembly bias, or reionization-feedback fronts, neighboring
halos can burst coherently on scales of order Mpc, adding a clustering-like contribution
to our shot noise. This effect is super-Poisson and can potentially bias the
inferred bursty enhancement upward from our per-halo calculation, most strongly at
large scales where it competes with the matter-driven clustering term. Furthermore, satellite galaxies hosted in the same halo could present correlated star formation processes, introducing an inter-halo temporal correlation. Taking this effect into account requires dedicated modeling. However, in the redshifts of interest for our work, satellite galaxies are predicted by simulations to constitute a moderate fraction of the total stellar mass of the Universe, and to have only a fairly small contribution to the total SFR~\cite{2023MNRAS.519.1578Y}. Neglecting them has little impact on the final results. 

Third, as mentioned in Sec.~\ref{sec:intro}, the formalism introduced here can be easily extended to MCGs, which have a non-negligible impact on the X-ray budget of the IGM~\cite{Lazare:2023jkg,Breitman:2026fte}. Such extension will in principle require introducing a separate burstiness amplitude  and correlation length for MCGs, as they generate stars via different physical mechanisms.  In order to propagate this burstiness model contribution to the 21-cm signal accurately, one would need either to model them based on first principles, or to calibrate against MCGs luminosity functions, in analogy to what has been done for ACGs in M26. As the data required for such calibration currently do not exist, we leave this study for future work.

Finally, in this study, we did not discuss the implications that a bursty SFR could have on the EoR, whose modeling is not included in the original \texttt{Zeus21} implementation. 
The mean ionizing photon production $\langle \dot
n_{\rm ion}\rangle$ is, by construction in our mean-anchored convention, unchanged by
burstiness. However, the ionizing photon
{field} inherits the OU statistics of $\dot M_\star$ on $\tauPS\!\sim\!25\,$Myr
timescales, with the same exponentially enhanced per-halo shot-noise contribution we
derive for Ly$\alpha$ and X-rays. This is expected to modify the bubble size distribution and the reionization patchiness in two ways. First, ionized bubbles around bursty galaxies, rather than growing monotonically, expand during bursts and may
contract slightly during quench phases through recombinations, with implications for the
21-cm power spectrum at $z\!\sim\!7$--$10$ that depend on the relative timescales
$t_{\rm rec}/\tauPS$. Such patchy-reionization signatures may carry
independent constraints on the burstiness-driven inhomogeneity in the ionizing-photon escape fraction $f_{\rm esc}$ and the
ionizing-photon production rate. Second, the relevant ionizing-photon production rate
becomes intrinsically stochastic at the few-Myr timescale of the UV-bright phase, so the
effective escape fraction $f_{\rm esc}$ is a function of burst phase and not just of
host-halo properties~\cite{Liu:2024fti,Sun:2023ewx}. A full treatment of these effects requires extending
the bursty shot-noise calculation to the ionizing-continuum field and propagating it
through a reionization morphology code; we defer this to a follow-up paper.

\section{Conclusions}\label{Sec:conclusions}
UVLF and $\rm H\alpha/UV$ data from JWST observations of the high-redshift universe have suggested that the early galaxy population was significantly  more variable than pre-launch expectations, prompting renewed attention to the role of bursty star formation in shaping early cosmic history. Building on this, the M26 analysis quantified the burstiness component of the SFR using an OU process model, and found a high level of intrinsic burstiness ($\sigma_{\rm PS}
\simeq 2.1$ on $\sim 25\,{\rm Myr}$ timescales). In this work we have folded that burstiness model into the
Zeus21 analytic 21-cm framework and investigated what the M26 result implies for the 21-cm power spectrum. 
We modeled the burstiness contribution as a single halo shot noise term in the SFRD power spectrum, and propagated it down to the radiation fields that drive the 21-cm signal. Our framework also includes a single-halo {\it deterministic} shot noise component which does not depend on burstiness, and should be included in the analytic computation in any case. 
We find that bursty stochasticity is not a small correction. Across a broad range of redshifts and scales, it boosts $\Delta^2_{T_{21}}$ by factors of several over the
{\it no-shot} baseline, with the largest enhancements appearing at the WF transition where the clustering signal cancels out, and on
small scales and at very high redshifts where the smooth signal is weak and is driven by small bursty halos. 
The impact of the {\it bursty-shot-noise} term depends on the chosen value of the unconstrained model parameter $\sigmax$ which caps the amplitude of the burstiness at the small halo-mass end. We show that the chosen value of this parameter mostly impacts the high redshift end, which is not accessible for most experiments planned for the upcoming decade, and that even a conservative choice that assumes no evolution of the amplitude at masses below the M26 data reach, still results in a significant enhancement to the signal at multiple scales and redshifts. 

Because the bursty boost preferentially populates halo masses below the JWST detection threshold, the 21-cm signal probes a
regime of the SFR distribution that is observationally inaccessible to JWST itself, and a joint analysis of the two datasets has the potential to pin down the
small-mass behavior of $\sigma_{\rm PS}(M_{\rm h})$ that a single probe struggles to determine on its own. 

This work paves the way to a more accurate analytical modeling of the 21-cm signal, which is essential to the study of 21-cm cosmology in the era of big data. 
Future work will quantify the resulting parameter degeneracies, with both cosmology and astrophysics, and will improve the modeling by including the timescales relevant for each of the fields that drive the 21-cm signal, and by extending this framework into the EoR.

Looking ahead, with a burst of data from 21-cm experiments expected in the coming decade, the 21-cm sky may soon reveal not only when the first galaxies formed their stars, but whether they did so in bursts.

\smallskip
\noindent\textit{Acknowledgments.}~%
HL is supported by the Zin PhD fellowship from the Kreitman School of Advanced Graduate
Studies at Ben-Gurion University. EV is supported by the Azrieli International Postdoctoral
Fellowship. JBM acknowledges support from NSF Grants AST-2307354 and AST-2408637, and NASA through grant JWST-GO-03224. EDK acknowledges support from the U.S.--Israel
Binational Science Foundation (NSF-BSF grant 2022743 and BSF grant
2024193) and the Israel National Science Foundation (ISF grant 3135/25),
as well as from the joint Israel--China program (ISF--NSFC grant
3156/23).


\clearpage
\appendix

\section{Detailed formalism}\label{app:detailed_form}

In the main text of the paper, we characterize burstiness through the scalar Gaussian random field $x(t)$; this is said to follow an Ornstein--Uhlenbeck (OU) process, namely it is the unique stationary solution of 
\begin{equation}
\dot x(t) \;=\; -\frac{1}{\tau_{\rm PS}}\,x(t) \;+\; \sqrt{\frac{2\sigma_x^{2}}{\tau_{\rm PS}}}\,\eta(t),
\label{eq:ou_langevin}
\end{equation}
where $\eta(t)$ is a unit-variance white noise. The two parameters $(\sigma_x^{2}, \tau_{\rm PS})$
are the stationary variance and the correlation time, and correspond to the quantities adopted in the main text. 

The SFR, however, is described by a lognormal distribution which depends on $e^{x(t)-\sigma_x^2/2}=y(t)$, where the term $\sigma_x^2/2$ is subtracted in our mean anchored convention. 
For the scope of the paper, we need therefore to characterize the statistical properties of $\langle e^{x(t_1)-\sigma_x^2/2}e^{x(t_2)-\sigma_x^2/2}\rangle$. 
To do so, we recall the properties of a distribution that combines two zero-mean Gaussian variables $X_{1}, X_{2}$ (jointly Gaussian, with arbitrary
covariance). The standard lognormal-moment identity is:
\begin{equation}
\begin{aligned}
\langle e^{X_{1}+X_{2}}\rangle &= \exp\!\left[\frac{{\rm Var}(X_{1}+X_{2})}{2}\,\right]\\
&=\exp\!\left[\frac{{\rm Var}(X_{1})}{2}+\frac{{\rm Var}(X_{2})}{2}+{\rm Cov}(X_{1},X_{2})\right].
\label{eq:lognormal_moment}
\end{aligned}
\end{equation}
This follows from the Gaussian moment generating function $\langle e^{a Y}\rangle\!=\!e^{a^{2}\sigma_{Y}^{2}/2}$
applied to $Y\!=\!X_{1}\!+\!X_{2}$. The covariance for the OU process, instead, is computed as 
\begin{equation}
{\rm Cov}(X_{1}, X_{2})\!=\!\langle x(t_{1})x(t_{2})\rangle\!-\!\langle x\rangle^{2}\!=\!\sigma_x^{2}e^{-|\Delta t|/\tau_{\rm PS}}.
\end{equation}
Substituting into Eq.~\eqref{eq:lognormal_moment}, we get:
\begin{align}
\langle y(t_{1})\,y(t_{2})\rangle
\;&=\; \langle e^{x(t_{1})+x(t_{2})}\rangle\,e^{-\sigma_x^{2}}\nonumber\\
\;&=\; \exp\!\left[\frac{{\rm Var}(x_{1}+x_{2})}{2}\right]\,e^{-\sigma_x^{2}}\nonumber\\
\;&=\; \exp\!\Bigl[\sigma_x^{2}\bigl(1 + e^{-|t_{1}-t_{2}|/\tau_{\rm PS}}\bigr)\Bigr]\,e^{-\sigma_x^{2}}\nonumber\\
\;&=\; \exp\!\Bigl[\sigma_x^{2}\,e^{-|t_{1}-t_{2}|/\tau_{\rm PS}}\Bigr],
\label{eq:y_two_point_app}
\end{align}
where the $-\sigma_x^{2}$ from the two factors of $e^{-\sigma_x^{2}/2}$ exactly cancels the
$+\sigma_x^{2}$ piece of the variance. This cancellation is the analytic content of the
mean-anchored convention. The remaining exponential is the OU correlator of $x(t)$ inside the exponential.

Finally, multiplying both sides of Eq.~\eqref{eq:y_two_point_app} by
$\overline{\dot{M}}_*(M_h, t_{1})\overline{\dot{M}}_*(M_h, t_{2})$ and using $\dot{M}_*\!=\!\overline{\dot{M}}_*\,y$ gives the
master result in Eq.~\eqref{eq:single_halo_master}.
The exponential of an
exponential is what makes the bursty correction nonlinear in $\sigma_x^{2}$: an
$\mathcal{O}(1)$ change in $\sigma_x^{2}$ at fixed $\Delta t=t_1-t_2$ produces an
exponentially-large change in the correlator, while an $\mathcal{O}(1)$ change in
$|\Delta t|/\tau_{\rm PS}$ at fixed $\sigma_x^{2}$ produces a smaller (logarithmic) change in the
inner exponent. 

\section{Effective model for the Cosmic Dawn 21-cm Signal}\label{sec:Zeus21}

The 21-cm signal is sourced by the spin-flip transition of neutral hydrogen (HI). Its brightness temperature, $T_{21}$, is defined relative to the CMB temperature, $T_{\rm CMB}$:\begin{equation}
    T_{21}(\boldsymbol{x},z)=\frac{T_s(\boldsymbol{x},z)-T_{\rm CMB}(\boldsymbol{x},z)}{1+z}(1-e^{-\tau_{21}(\boldsymbol{x},z)}),
\end{equation}
where $T_s$ is the spin temperature,~i.e.~the population ratio between HI in the triplet and singlet state. Instead, the 21-cm optical depth is 
\begin{equation}
    \tau_{21}=(1+\delta_b)x_{\rm HI}\frac{T_0}{T_s}\frac{H(z)}{\partial_r v_r}(1+z),
\end{equation}
$\delta_b$ being the baryon density, $x_{\rm HI}$ the neutral hydrogen fraction, $H(z) $ the Hubble expansion rate, $\partial_r v_r$ the line-of-sight (LoS) velocity gradient and $T_0$ a cosmology-dependent normalization factor. 

Ref.~\cite{Munoz:2023kkg} relied on this fact to build a fully analytical model to estimate the 21-cm power spectrum during Cosmic Dawn, released with the public code \texttt{Zeus21}. This was then extended in Ref.~\cite{Cruz:2024fsv} to include population III stars and Ref.~\cite{Libanore:2025wtu} to consistently model the emission of other lines, associated with star formation ([OIII], [OII], H$\alpha$, H$\beta$, CO, [CII]..., implemented in the public code \texttt{oLIMpus}). A new version of \texttt{Zeus21} will soon be released~\cite{Sklansky:inprep}, extending the computation of the 21-cm power spectrum toward the EoR. 

The core assumption beneath all these codes is that the SFRD can be approximated as a lognormal function of the smoothed density field $\delta_R$,
\begin{equation}
\begin{aligned}
    \dot{\rho}_*(z|\delta_R)&=\overline{\dot{\rho}}_*(z)e^{\gamma_R\tilde{\delta}_R}\\
    &=(1+\delta_R)\int dM_h \frac{dn_{\rm EPS}(\delta_R,z)}{dM_h}\dot{M}_*(M_h,z),
    \label{eq:smoothed_SFRD}
\end{aligned}
\end{equation}
where $\overline{\dot{\rho}}_*(z)$ is the average SFRD, and $\tilde{\delta}_R=\delta_R-\gamma_R\sigma_R^2/2$ for normalization purposes, with $\sigma_R^2$ the variance of the smoothed density field. The coefficient $\gamma_R$ is calibrated from the second line of Eq.~\eqref{eq:smoothed_SFRD}, in which the SFRD is estimated by integrating over halo masses $M_h$ the product between the SFR ($\dot{M}_*$) and the halo mass function (HMF, $dn/dM_h$). This, in the extended Press-Schechter formalism (EPS), can be computed as a function of $\delta_R$. 

We compute the SFR $\dot M_*$ as $\dot M_* = f_b\cdot f_*\cdot f_{\rm duty} \cdot \dot{M_h}$, where $f_b$ is the baryonic fraction $f_b = \Omega_b/\Omega_m$ ($\Omega_{b,m}$ the cosmological baryon and total-matter density parameters, respectively) and $f_*$ is the star formation efficiency, given by the empirical relation
\begin{equation}\label{eq:SHMR}
    f_*(M_h,z) = \frac{2\epsilon_*}{(M_h/M_p)^{\alpha_\ast} + (M_h/M_p)^{\beta_\ast}},
\end{equation}
with $\alpha_\ast > 0$
 and $\beta_\ast <0$. This expression is inspired by the observed stellar-to-halo mass ratio (SHMR), which peaks at halo mass scale $M_p$, and decays both for smaller and larger halos, due to supernova and photo-ionization feedback and AGN feedback respectively.
Finally, the duty cycle $f_{\rm duty}$ exponentially cuts off galaxy formation in very small halos, while the halo accretion rate $\dot{M_h}$ follows an exponential halo growth model $M_h(z) \propto M_h \exp(-\alpha z)$ where $\alpha = 0.79$ from simulations~\cite{Dekel:2013uaa}. 
Whenever referring to the average SFR, $\overline{\dot{M}}_*$ in the main text, this is computed relying on this model.

\subsection{21-cm clustering power spectrum}

Under the approximation in Eq.~\eqref{eq:smoothed_SFRD}, one can compute the SFRD real-space two-point function as~\cite{Coles:1991if,Xavier:2016elr} 
\begin{equation}
    \langle e^{\gamma_{R_1}\tilde{\delta}_{R_1}}e^{\gamma_{R_2}\tilde{\delta}_{R_2}}\rangle=e^{\gamma_{R_1}\gamma_{R_2}\xi_{R_1R_2}}-1,
\label{eq:SFRD_xi}
\end{equation}
where $\xi_{R_1R_2}={\rm FT}[P_{mm}(k)W^{\rm TH}_{R_1}(k)W^{\rm TH}_{R_2}(k)]$ is the Fourier transform of the matter power spectrum, smoothed using the real-space top-hat window function $W^{\rm TH}_{R}=3[\cos(kR)-(kR)\sin(kR)]/(kR)^3$.

Eqs.~\eqref{eq:smoothed_SFRD} and~\eqref{eq:SFRD_xi} are the main building blocks of the 21-cm power-spectrum computation in \texttt{Zeus21}. In fact, substituting the approximated lognormal SFRD in the flux computation in Eq.~\eqref{eq:fluxes_general}, the two-point functions of the radiative fields are obtained by integrating over concentric shells $R_1,R_2$. The Ly$\alpha$ auto-correlation can be expressed as 
\begin{equation}
\begin{aligned}
    \xi_{\rm Ly\alpha}(z)&=c_{1,\alpha}^2(z)\sum_{R_1,R_2}c_{2,\alpha}(R_1,z(R_1))c_{2,\alpha}(R_2,z(R_2))\\
    &\quad\times (e^{\gamma_{R_1}\gamma_{R_2}\xi_{R_1R_2}}-1),
\end{aligned}
\label{eq:Lyalpha_xi_clustering}
\end{equation}
where we substituted $\int dR\to \sum_R\Delta R$, we defined $z(R)$ as the redshift corresponding to the shell at comoving distance $R$, and the coefficients are 
\begin{align}
    c_{1,\alpha}(z)&=\frac{(1+z)^2}{4\pi}\frac{S_\alpha}{J_{\alpha,c}},\\
    c_{2,\alpha}(R,z(R))&=\Delta R\,\epsilon_\alpha(R)\,\overline{\dot{\rho}}_*(z(R)).
\end{align}
In the first line, the factor $J_{\alpha,c}$ is a normalization constant, while $S_\alpha$ is the correction factor from Ref.~\cite{Hirata:2005mz}. In the second line, the quantity $\epsilon_\alpha(R)$ represents the total-Ly$\alpha$ spectral energy distribution (SED), namely the sum over all transitions that eventually shift into Ly$\alpha$ at $z$. 

Similarly, we can also express the X-ray auto-correlation; in this case, however, we need to take into account the fact that the temperature of the gas in $(\boldsymbol{x},z)$ depends on the gas heating rate integrated over the past light cone; therefore, the X-ray two-point function acquires an additional integral over redshifts,
\begin{equation}
\begin{aligned}
    \xi_{T_X}&=\sum_{z'_1,z'_2\geq z}c_{1,X}(z_1')c_{1,X}(z_2')\sum_{R_1,R_2}c_{2,X}(R_1,z_1'(R_1))\\
    &\quad\times c_{2,X}(R_2,z_2'(R_2))\bigl(e^{\gamma_{R_1}\gamma_{R_2}\xi_{R_1R_2}}-1\bigr),
\end{aligned}
\label{eq:Xray_xi_clustering}
\end{equation}

where, similarly to the previous case, $c_{1,X}$ contains information on the redshift evolution of the radiation field and its normalization, while $c_{2,X}(R,z(R))=\Delta R\,\epsilon_X(R,z)e^{-\tau_X(z)}\overline{\dot{\rho}}_*(z)$ with $\epsilon_X$ the X-ray SED, and $\tau_X$ the X-ray opacity term~\cite{Pritchard:2006sq}.

Finally, Eq.~\eqref{eq:Lyalpha_xi_clustering} and~\eqref{eq:Xray_xi_clustering} can be Fourier transformed to estimate the contribution of Ly$\alpha$ and X-rays to the 21-cm {\it clustering} power spectrum. As Ref.~\cite{Munoz:2023kkg} explains in detail, computing the 21-cm power spectrum also requires calculating the cross-power spectrum between Ly$\alpha$ and X-rays, as well as the contribution of the density power spectrum and of its cross-correlation with the two radiation fields. Eventually, the 21-cm power spectrum is computed as 
\begin{equation}
    \Delta^2_{21}=\sum_{i,j}\beta_i\beta_j\Delta^2_{ij},
    \label{eq:power_21cm}
\end{equation}
where $i,j=\{{\rm Ly\alpha},T_X,\delta\}$, and the coefficients of the linear combination $\beta_i$ are estimated from the partial derivative of $T_{21}$ with respect to each component.

\end{document}